\documentclass[aip,reprint]{revtex4-1}

\usepackage{graphicx}
\usepackage{amsmath}
\usepackage{amssymb}

\usepackage{color}

\makeatletter
\renewcommand{\p@subsection}{}
\renewcommand{\p@subsubsection}{}
\makeatother

\draft 
\begin{document}

\title{Dynamics from seconds to hours in Hodgkin--Huxley model with time--dependent ion concentrations and buffer reservoirs} 

\author{Niklas H\"ubel}
\affiliation{Department of Theoretical Physics, Technische Universit{\"a}t Berlin, Germany}

\author{Markus A. Dahlem}
\thanks{Corresponding author}
\email[]{markus.dahlem@gmail.com\\Current address:\\
Department Biological Physics\\Max Planck Institute for the Physics of Complex Systems\\
N\"othnitzer Straße 38\\
01187 Dresden\\
Germany
}
\affiliation{Department of Physics, Humboldt Universit\"at zu Berlin, Berlin, Germany}

\date{\today}

\begin{abstract}

The classical Hodgkin--Huxley (HH) model neglects the time--dependence of ion
concentrations in spiking dynamics. The dynamics is therefore limited to a time
scale of milliseconds, which is determined by the membrane capacitance
multiplied by the resistance of the ion channels, and by the gating time
constants.  We study slow dynamics in an extended HH framework that includes
time--dependent ion  concentrations, pumps, and buffers. Fluxes across the
neuronal membrane change intra-- and extracellular ion concentrations, whereby
the latter can also change through contact to reservoirs in the surroundings.
Ion gain and loss of the system is identified as a bifurcation parameter whose
essential  importance was not realized in earlier studies. Our systematic study
of the bifurcation structure and thus the phase space structure helps to understand
activation and inhibition of a new excitability in ion  homeostasis which
emerges in such extended models. Also modulatory mechanisms that  regulate the
spiking rate can be explained by bifurcations. The dynamics on three distinct
slow times scales is determined by the cell volume--to--surface--area ratio and
the membrane permeability (seconds), the buffer time constants (tens of
seconds), and the slower backward buffering (minutes to hours). The modulatory
dynamics and the newly emerging excitable dynamics corresponds to pathological  
conditions observed in epileptiform burst activity, and spreading depression in 
migraine aura and stroke, respectively. 

\end{abstract}

\maketitle

\section*{Author summary}\label{sec:0}
The classical theory by Hodgkin and Huxley (HH) describes nerve impulses
(spikes) that manifest communication between nerve cells. The underlying
mechanism of a single spike is excitability, i.e.,  a small disturbance
triggers a large excursion that reverts without further input to the original state.  A
spike lasts a 1/1000 second and even though during this period ions are
exchanged across the nerve cell membrane, the change in the corresponding ion
concentrations can become significant only in series of such spikes. Under
certain pathological conditions changes in ion concentrations become massive
and last minutes to hours before they recover. This establishes a new type of
excitability underlying communication failure between nerve cells during
migraine and stroke. To clarify this mechanism and to recognize the relevant
factors that determine the slow time scales of ion changes, we use an extended
version of the classical HH theory. We identify one variable of particular
importance, the potassium ion gain or loss through some reservoirs provided by
the nerve cell surroundings. We suggest to describe the new excitability as a
sequence of two fast processes with constant total ion content separated by two
slow processes of ion clearance (loss) and re--uptake (re--gain).

\section*{Introduction}\label{sec:1} 
In this paper we study ion dynamics in ion--based neuron models. In comparison to 
classical HH type membrane models this introduces dynamics on much slower time scales. 
While spiking activity is in the order of milliseconds, the time scales of ion 
dynamics range from seconds to minutes and even hours depending on the process 
(transmembrane fluxes, glial buffering, backward buffering). The slow dynamics leads 
to new phenomena. Slow burst modulation as in seizure--like activity (SLA) emerges from 
moderate changes in the ion concentrations. Phase space excursions with large changes 
in the ionic variables establish a new type of ionic excitability as observed in 
cortical spreading depression (SD) during stroke and in migraine with aura 
\cite{DRE11,CHA13a}. Such newly emerging dynamics can be understood from the phase 
space structure of the ion--based models. 

Mathematical models of neural ion dynamics can be divided into two classes. On the one 
hand the discovery of SD by Le\~ao in 1944 \cite{LEA44}---a severe perturbation of 
neural ion homeostasis associated with a huge changes in the potassium, sodium and 
chloride ion concentrations in the extracellular space (ECS)\cite{MAR00a} that spreads 
through the tissue---has attracted many modelling approaches dealing with the 
propagation of large ion concentration variations in tissue. In 1963 Grafstein described 
spatial potassium dynamics 
during SD in a reaction--diffusion framework with a phenomenological cubic rate 
function for the local potassium release by the neurons\cite{GRA63}. Reshodko and 
Bur\'es proposed an even simpler cellular automata model for SD propagation\cite{RES75}. 
In 1978 Tuckwell and Miura developed a SD model that is amenable to a more direct 
interpretation in terms of biophysical quantities \cite{TUC78}. It contains ion 
movements across the neural membrane and ion diffusion in the ECS. In more recent 
studies Dahlem et al.\ suggested certain refinements of the spatial coupling mechanisms, 
e.g., the inclusion of nonlocal and time--delayed feedback terms to explain very 
specific patterns of SD propagation in pathological situations like migraine with 
aura and stroke \cite{DAH09a,DAH12b}.

On the other hand single cell ion dynamics were studied in HH--like membrane models 
that were extended to include ion changes in the intracellular space (ICS) and the ECS 
since the 1980s. While the first extensions of this type were developed for cardiac 
cells by DiFranceso and Noble\cite{DIF85,DOK93}, the first cortical model in this 
spirit was developed by Kager, Wadman and Somjen (KWS)\cite{KAG00} only in 2000. Their 
model contains abundant physiological detail in terms of morphology and ion channels, 
and was in fact designed for seizure--like activity (SLA) and local SD dynamics. It 
succeeded spectacularly in reproducing the experimentally known phenomenology. An even 
more detailed model was proposed by Shapiro at the same time\cite{SHA01} who---like Yao, 
Huang and Miura for KWS\cite{YAO11}---also investigated SD propagation with a spatial 
continuum ansatz. 

In the following HH--like models of intermediate complexity were developed by 
Fr\"ohlich, Bazhenov et al.\ to describe potassium dynamics during epileptiform 
bursting\cite{FRO06,FRO08,BAZ04}. The simplest HH--like model of cortical ion 
dynamics was developed by Barreto, Cressman et al.\cite{CRE09,BAR11} who describe the 
effect of ion dynamics in epileptiform bursting modulation in a single compartment 
model that is based on the classical HH ion channels. Interestingly in none of these 
considerably simpler than Shapiro and KWS models extreme ion dynamics like in SD or 
stroke was studied. To our knowledge the only exception is a study by Zandt et al.\ 
who describe in the framework of Cressman et al.\ what they call the ``wave of death'' 
that follows the anoxic depolarization after decapitation as measured in experiments 
with rats\cite{ZAN11}.

\begin{figure}[t!]
\begin{center}
\includegraphics[width=0.975\columnwidth]{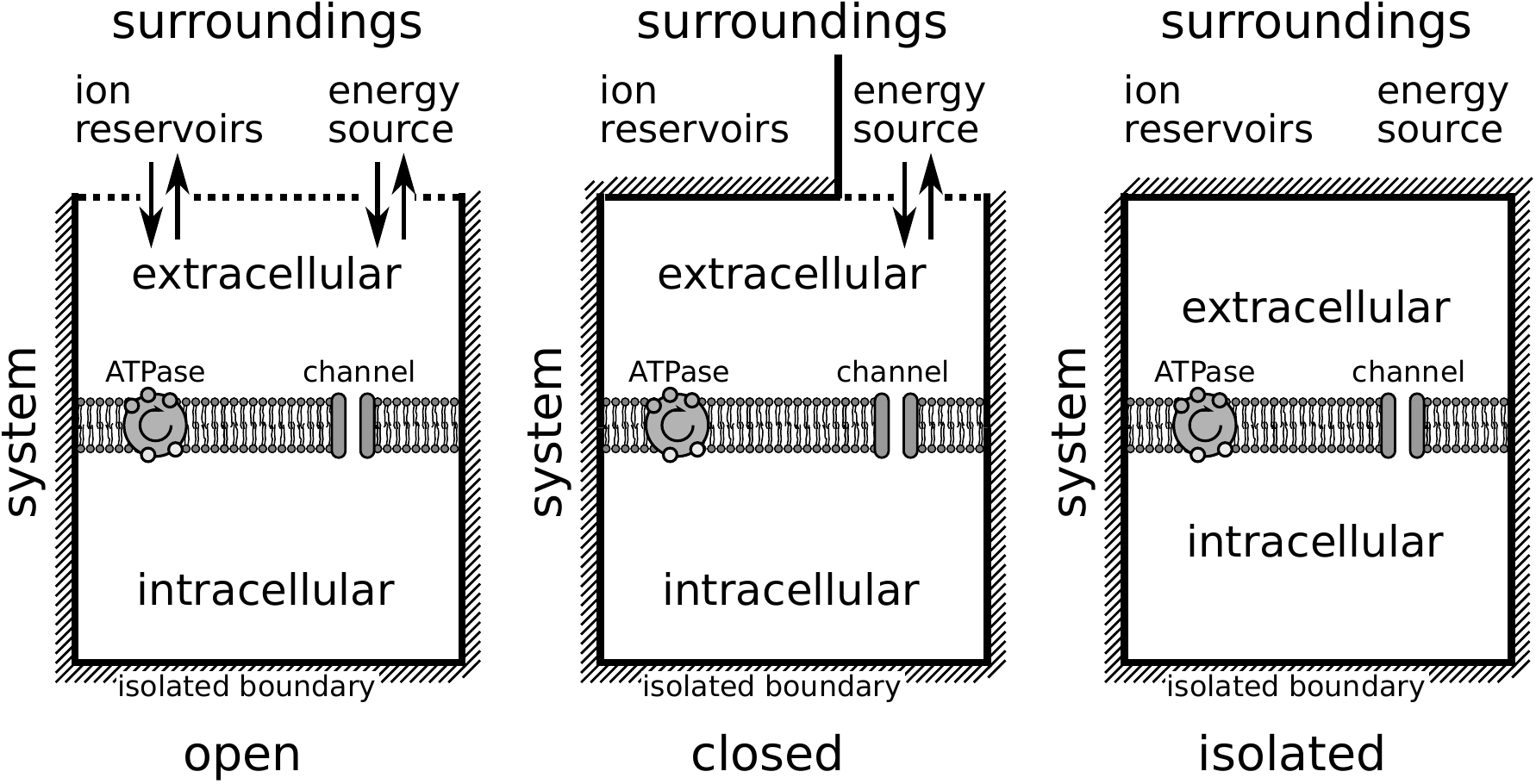}
\end{center}\caption{The ion--based model describes a heterogeneous system, comprising 
extracellular and intracellular compartments separated by a membrane, and the 
surroundings of the system. The latter provides an energy source and, if the system is 
not closed, also an ion reservoir. \label{fig:system}}
\end{figure}

In this study we systematically analyze the entire phase space of such local 
ion--based neuron models containing the full dynamical repertoire ranging from fast 
action potentials to slow changes in ion concentrations. We start with the simplest 
possible model for SD dynamics---a variation of the Barreto, Cressman et al.\ 
model---and reproduce most of the results for the KWS model. Our analysis covers SLA 
and SD.

Three situations should be distinguished: isolated, closed, and open systems, which is 
reminiscent of a thermodynamic viewpoint (see Fig.~\ref{fig:system}). An isolated system 
without transfer of metabolic energy for the ATPase--driven $\text{Na}^+/\text{K}^+$ 
pumps will attain its thermodynamic equilibrium, i.e., its Donnan equilibrium. A closed 
neuron system with functioning pumps but without ion regulation by glia cells or the 
vascular system is generally bistable\cite{HUE14}. There is a stable state of free 
energy--starvation (FES) that is close to the Donnan equilibrium and coexists with 
the physiological resting state. The ion pumps cannot recover the physiological 
resting state from FES.

We will now develop a novel phase space perspective on the dynamics in open neuron 
systems. We describe the first slow--fast decomposition of local SD dynamics, in 
which the ion gain and loss through external reservoirs is identified as the crucial 
quantity whose essential importance was not realized in earlier studies. Treating 
this slow variable as a parameter allows us to derive thresholds for SD ignition and 
the abrupt, subsequent repolarization of the membrane in a bifurcation analysis for 
the first time. Moreover we analyze oscillatory dynamics in open systems and thereby 
relate SLA and SD to different so--called torus bifurcations. This categorizes SLA 
and SD as genuinely different though they are `sibling' dynamics as they both bifurcate 
from the same `parent' limit cycle in a supercritical and subcritical manner, respectively, 
which also explains the all--or--none nature of SD. SLA is gradual in contrast.

\section*{Model}\label{sec:2}
Local ion dynamics of neurons has been studied in models of various complexity. Reduced model 
types consist of an electrically excitable membrane containing gated ion channels and 
ion concentrations in an intra-- and an extracellular compartment \cite{CRE09,ZAN11,BAR11}. 
Transmembrane currents must be converted to ion fluxes that lead to changes in the 
compartmental ion concentrations. Such an extension requires ion pumps to prevent the 
differences between ICS and ECS ion concentrations that are present under physiological 
resting conditions from depleting.

We consider a model containing sodium, potassium and chloride ions. The 
HH--like membrane dynamics is described by the membrane potential $V$ and
the potassium activation variable $n$. The sodium activation $m$ is approximated 
adiabatically and the sodium inactivation $h$ follows from an assumed functional 
relation between $h$ and $n$. The ICS and ECS concentrations of sodium, 
potassium and chloride ions are denoted by $\mathit{Na}_{i/e}$, $K_{i/e}$ and 
$\mathit{Cl}_{i/e}$, respectively. 

In a closed system mass conservation holds, i.e.,
\begin{eqnarray}
\mathit{ion}_i\cdot\omega_i + \mathit{ion}_e\cdot\omega_e = \mathit{const.}
\end{eqnarray}
with $\mathit{ion}\in \{\textit{Na}^+,\ K^+,\ \textit{Cl}^-\}$ and the ICS/ECS volumes 
$\omega_{i/e}$. Together with the electroneutrality of ion fluxes across the membrane, i.e., 
\begin{eqnarray}
Q_i:=K_i+\mathit{Na}_i - \mathit{Cl}_i = \mathit{const.}\ ,\label{eq:0}
\end{eqnarray}
only two of the six ion concentrations are independent dynamical variables. The full list 
of rate equations then reads
\begin{eqnarray}
\frac{\mathrm{d}V}{\mathrm{d}t} 			&=& -\frac{1}{C_m}(I_{Na^+}+I_{K^+} + I_{Cl^-} + I_p)\ ,	\label{eq:1}\\
\frac{\mathrm{d}n}{\mathrm{d}t} 			&=&  \phi\frac{n_\infty -n}{\tau_n}\ ,						\label{eq:2}\\
\frac{\mathrm{d}K_i}{\mathrm{d}t}			&=&	-\frac{\gamma}{\omega_i}(I_{K^+}-2 I_p)\ ,				\label{eq:3}\\
\frac{\mathrm{d}\mathit{Cl}_i}{\mathrm{d}t}	&=&	+\frac{\gamma}{\omega_i}I_{Cl^-}\ .						\label{eq:4}
\end{eqnarray}

They are complemented by six constraints on gating variables and ion concentrations:
\begin{eqnarray}
\mathit{Na}_i	&=&\mathit{Na}_i^0+(K_i^0-K_i)-(\mathit{Cl}_i^0-\mathit{Cl}_i)					\ ,\label{eq:5}\\
\mathit{Na}_e	&=&\mathit{Na}_e^0+\frac{\omega_i}{\omega_e}(\mathit{Na}_i^0	-\mathit{Na}_i)	\ ,\label{eq:6}\\
\mathit{K}_e	&=&\mathit{K}_e^0	+\frac{\omega_i}{\omega_e}(\mathit{K}_i^0	-\mathit{K}_i)	\ ,\label{eq:7}\\
\mathit{Cl}_e	&=&\mathit{Cl}_e^0+\frac{\omega_i}{\omega_e}(\mathit{Cl}_i^0	-\mathit{Cl}_i)	\ ,\label{eq:8}\\
m				&=& m_\infty																	\ ,\label{eq:9}\\
h				&=& 1 - \frac{1}{1 + \exp(-6{.}5(n-0{.}35))}									\ .\label{eq:10}
\end{eqnarray}

Superscript 0 of indicates ion concentrations in the physiological resting state. 
Unless otherwise stated $K_i^0$ and $\mathit{Cl}_i^0$ are used as initial conditions in 
the simulations. Constrained ion concentrations (Eqs.~(\ref{eq:5})--(\ref{eq:8})) then 
also take their physiological resting state values. These ion concentrations, the 
membrane capacitance $C_m$, the gating time scale parameter $\phi$, the conversion 
factor $\gamma$ from currents to ion fluxes, and the ICS and ECS volumes $\omega_{i/e}$ 
are listed in Tab.~\ref{tab:1}. The conversion factor $\gamma$ is an expression of the 
membrane surface area $A_m$ and Faraday's constant $F$ (both given in Tab.~\ref{tab:1}, too):
\begin{eqnarray}
\gamma = \frac{A_m}{F}\ \label{eq:11}
\end{eqnarray}
We remark that all parameters in Tab.~\ref{tab:1} are given in typical units of the 
respective quantities. The numerical values in these units can directly be used for 
simulations. Time is then given in msec, the membrane potential in mV and ion 
concentrations in $\mathrm{mM}$. 

The electroneutrality of the total transmembrane ion flux as expressed in 
Eqs.~(\ref{eq:0}) and (\ref{eq:5}) is a consequence of the large time scale separation 
between the membrane dynamics and the ion dynamics (cf.\ Ref.\ \cite{HUE14} and the 
below discussion of time scales). This constraint is the reason why the thermodynamic 
equilibrium of the system must be understood as a Donnan equilibrium. This is the 
electrochemical equilibrium of a system with a membrane that is impermeable to some 
charged particles, which can be reached in an electroneutral fashion, i.e., without separating 
charges. We do not include this impermeant matter explicitly, because it does not influence 
the dynamics as long as osmosis is not considered. One should however keep in mind that 
the initial ion concentrations in Tab.~\ref{tab:1} do not imply zero charge in the ICS or ECS 
and hence impermeant matter to compensate for this must be present.

The gating functions $n_\infty$, $\tau_n$ and $m_\infty$ are given by
\begin{eqnarray}
n_\infty &=& \frac{\alpha_n}{\alpha_n + \beta_n}	\ ,\label{eq:12}\\
\tau_n   &=& \frac{1}{\phi(\alpha_n+\beta_n)}		\ ,\label{eq:13}\\
m_\infty &=& \frac{\alpha_m}{\alpha_m + \beta_m}	\ .\label{eq:14}
\end{eqnarray}
Here $n_\infty$ and $m_\infty$ are the asymptotic values and $\tau_n$ is potassium 
activation time scale. They are expressed in terms of the Hodgkin--Huxley exponential 
functions\cite{BAR11,CRE09}
 \begin{eqnarray}
\alpha_m	&=& \frac{0{.}1(V+30)}{1 - \exp(-(V+30)/10)}	\ ,\label{eq:15}\\
\beta_m		&=& 4 \exp(-(V+55)/18)							\ ,\label{eq:16}\\
\alpha_n	&=& \frac{0{.}01(V+34)}{1 - \exp(-(V+34)/10)}	\ ,\label{eq:17}\\
\beta_n		&=& 0{.}125 \exp(-(V+44)/80) 	 				\ .\label{eq:18}
\end{eqnarray}

The three ion currents are 
\begin{eqnarray}
I_{Na^+} &=& (g_{Na}^l + g_{Na}^g m^3h) \cdot (V - E_{Na})\ ,	\label{eq:19}\\
I_{K^+}  &=& (g_K^l + g_K^g n^4) \cdot (V - E_K)\ ,				\label{eq:20}\\
I_{Cl^-} &=&  g_{Cl}^l \cdot (V - E_{Cl})\ .					\label{eq:21}
\end{eqnarray}
They are given in terms of the leak and gated conductances $g_{ion}^{l,g}$ (with 
$\mathit{ion}\in \{\mathit{Na}^+,\ K^+,\ \mathit{Cl}^-\}$) and the Nernst potentials 
$E_{ion}$ which are computed from the (dynamical) ion concentrations 
$\mathit{ion}_{i/e}$:
\begin{eqnarray}
E_{ion}=\frac{26{.}64}{z_{ion}}\ln(\mathit{ion}_e/\mathit{ion}_i)\ ,\label{eq:22}
\end{eqnarray}
$z_{ion}$ denotes the valence of the particular ion species. 

The pump current modelling the ATPase--driven exchange of intracellular sodium with
extracellular potassium at a $3/2$--ratio is given by
\begin{eqnarray}
I_p(\mathit{Na}_i,K_e)&=&\rho\bigg(1+\exp{\bigg(\frac{25-\mathit{Na}_i}{3}\bigg)}\bigg)^{-1} \nonumber \\ 
&& \bigg(1+\exp{(5.5-K_e)}\bigg)^{-1}\ ,\label{eq:23}
\end{eqnarray}
where $\rho$ is the maximal pump rate\cite{BAR11}. The pump current increases with
$\mathit{Na}_i$ and $K_e$. The values for the conductances and pump rate are
also given in Tab.~\ref{tab:1}. Let us remark that in comparisons with Ref.\
\cite{HUE14}, we have mildly increased the maximal pump rate and decreased the
chloride conductance to obtain a SD threshold in agreement with experiments
(see Sect.\ Results).

Eqs.~(\ref{eq:1})--(\ref{eq:10}) describe a closed system in which ion pumps
are the only mechanism maintaining ion homeostasis and in which mass
conservation holds for each ion species. A remark on terminology is due at this
point: a `closed' system refers exclusively to the conservation of the ion
species that we model.  We do not directly model other mass transfer that
occurs in real neural systems.  Yet it is indirectly included. The ion pumps
use energy released by hydrolysis of ATP, a molecule whose
components (glucose and oxygen or lactate) therefore have to pass the
system boundaries.  In thermodynamics, it is customary to call systems that
exchange energy but not matter with their environment closed. Since ATP is in
this framework only considered as an energy source, we can describe the system
as closed, if ions cannot be transferred across its boundaries.

As mentioned above the closed system is bistable. Superthreshold stimulations 
cause a transition from physiological resting conditions to FES. To resolve this 
and change the behaviour to local SD dynamics it is necessary to include further 
regulation mechanisms\cite{HUE14}. Since SD is in particular characterized by an 
extreme elevation of potassium in the ECS we will only discuss potassium regulation. 

If ECS potassium ions are subject to a regulation mechanism which is
independent of the membrane dynamics, then the symmetry between ICS
and ECS potassium dynamics is broken and Eq.~(\ref{eq:7}) for the
potassium conservation does not hold. Let us represent changes of the
potassium content of the system by a variable $\tilde{K}_e$ which is defined by
the following relation: 
\begin{eqnarray}
\mathit{K}_e=\mathit{K}_e^0+\frac{\omega_i}{\omega_e}(\mathit{K}_i^0-\mathit{K}_i)+\tilde{K}_e\ \label{eq:24} 
\end{eqnarray}

Changes of the potassium content, i.e., changes of $\tilde{K}_e$, can be of different 
physiological origin. If glial buffering is at work the potassium content will be 
reduced by the amount of buffered potassium $K_b$. An initial potassium elevation 
$\Delta K_e^0$ simply leads to an accordingly increased $\tilde{K}_e$:
\begin{eqnarray}
\tilde{K}_e=\Delta K_e^0 - K_b\ . \label{eq:25}
\end{eqnarray}
For the coupling to an extracellular potassium bath or to the vasculature $\tilde{K}_e$ 
is a measure for the amount of potassium that has diffused into (positive $\tilde{K}_e$) 
or out of (negative $\tilde{K}_e$) the system.

We are going to discuss two regulation schemes---coupling to an extracellular bath and 
glial buffering. They could be implemented simultaneously, but for our purpose it will 
suffice to apply only one scheme at a time. In the second subsection of Sect.\ Results, the dynamics of 
$\tilde{K}_e$ is given by glial buffering, while in the third subsection we will discuss the 
oscillatory regimes one finds for bath coupling with elevated bath concentrations. To 
implement glial buffering we assume a phenomenological chemical reaction of the 
following type\cite{KAG00,CHA13}:
\begin{eqnarray}
K_e+B\underset{k_1}{\overset{k_2}{\rightleftharpoons}} K_b \label{eq:26}
\end{eqnarray}
The buffer concentration is denoted by $B$. We are using the buffer model from 
Ref.\ \cite{KAG00} in which the potassium--dependent buffering rate $k_2$ is given as
\begin{eqnarray}
k_2 = \frac{\bar{k}_1}{1+\exp(-(K_e-15)/1{.}09)}\ . \label{eq:27}
\end{eqnarray}
The parameter $\bar{k}_1$ is normally assumed to have the same numerical value as the 
constant backward buffering rate $k_1$ which is hence an overall parameter for the 
buffering strength. However, the parameters should be denoted differently as they have 
different units (cf.\ Tab.~\ref{tab:1}). This chemical reaction scheme together with the 
mass conservation constraint
\begin{eqnarray}
B^0 = K_b + B\ , \label{eq:28}
\end{eqnarray}
where $B^0$ is the initial buffer concentration, leads to the following differential 
equation for $K_b$:
\begin{eqnarray}
\frac{\mathrm{d}K_b}{\mathrm{d}t} = k_2 K_e (B_0 - K_b) - k_1K_b  \label{eq:29}
\end{eqnarray}
Eq.~(\ref{eq:25}) the implies the following rate equation for $\tilde{K}_e$
\begin{eqnarray}
\frac{\mathrm{d}\tilde{K}_e}{\mathrm{d}t} = -k_2 K_e (B_0 - K_b) + k_1K_b\ . \label{eq:30}
\end{eqnarray}
where $K_b$ and $K_e$ are given by Eqs.~(\ref{eq:25}) and (\ref{eq:24}), respectively.

To model the coupling to a potassium bath one normally includes an explicit rate 
equation for the ECS potassium concentration
\begin{eqnarray}
\frac{\mathrm{d}K_e}{\mathrm{d}t} = -\frac{\omega_i}{\omega_e}\frac{\mathrm{d}K_i}{\mathrm{d}t} + J_{\textit{diff}}\ , \label{eq:31}
\end{eqnarray}
where the diffusive coupling flux 
\begin{eqnarray}
J_{\mathit{diff}} = \lambda(K_\mathit{bath}-K_e)\ . \label{eq:32}
\end{eqnarray}
is defined by its coupling strength $\lambda$ and the potassium bath concentration 
$K_{bath}$. Eq.~(\ref{eq:24}) implies that Eq.~(\ref{eq:31}) can be rewritten in terms 
of $\tilde{K}_e$ as follows:
\begin{eqnarray}
\frac{\mathrm{d}\tilde{K}_e}{\mathrm{d}t} = J_{\textit{diff}} \label{eq:33}
\end{eqnarray}
Note that we have chosen to formulate ion regulation in terms of $\tilde{K}_e$ rather 
than $K_e$ which would be completely equivalent. This is crucial, because the dynamics of 
$\tilde{K}_e$ happens on a time scale that is only defined by the buffering or the 
diffusive process, while $K_e$ dynamics involves transmembrane fluxes and reservoir coupling 
dynamics at different time scales (cf.\ the last paragraph of this section). This can be 
seen from Eq.~(\ref{eq:31}).

Both regulation schemes---glial buffering given by Eq.~(\ref{eq:30}) and coupling to a 
bath with a physiological bath concentration as in Eq.~(\ref{eq:33})---can be used to 
change the system dynamics from bistable to ionically excitable, i.e., excitable with 
large excursions in the ionic variables. Like all other system parameters the 
regulation parameters $k_1$ and $\lambda$ are given in Tab.~\ref{tab:1}. They have been 
adjusted so that the duration of the depolarized phase is in agreement with 
experimental data on spreading depression. 

Note that the parameters we have chosen are up to almost one order of magnitude lower 
than intact brain values like the ones used in Refs.\ \cite{ULL10,KAG00,CHA13}. 
While this does not affect the general time scale separation between glial or vascular 
ion regulation and ion fluxes across the cellular membrane, the duration of SD depends 
crucially on these parameters. However, during SD oxygen deprivation will weaken glial
buffering, and the swelling of glia cells and blood vessel constriction will restrict 
diffusion to the vasculature. Such processes can be included to ion--based neuron models and make 
ion regulation during SD much slower\cite{ULL10,KAG00,CHA13}. 

For our purpose it is however sufficient to assume smaller values from the beginning. 
We remark that the ion regulation schemes in our model only refer to vascular coupling 
and glial buffering. Lateral ion movement between the ECS of nearby neurons is a different 
diffusive process that determines the velocity of a travelling SD wave in tissue. This is 
not described in our framework. In the following section we will demonstrate in 
detail how $\tilde{K}_e$ can be understood as the inhibitory variable of this excitation 
process.

The above presented model is indeed the simplest ion--based neuron model that exhibits 
local SD dynamics. Model simplicity is an appealing feature in its own right, but one might
doubt the physiological relevance of such a reduced model. Our hypothesis is that it 
captures very general dynamical features of neuronal ion dynamics, and to confirm this we 
will compare the results obtained with the reduced model to the physiologically much more 
detailed KWS model\cite{KAG00}. This detailed model contains five different gated ion 
channels (transient and persistent sodium, delayed rectifier and transient potassium, and 
NMDA receptor gated currents) and has been used intensively to study SD and SLA. In fact, 
one modification is required so that we can replicate the results obtained from the reduced 
model. The KWS model contains an unphysical so--called 'fixed leak' current
\begin{eqnarray}
I_\mathit{leak,f} = g_\mathit{leak,f} \cdot (V+70) \label{eq:34}
\end{eqnarray}
that has a constant reversal potential of $-70$ mV and no associated ion species.
This current only enters the rate equation for the membrane potential $V$. 

The effect on the model dynamics is dramatic. To see this note that the electroneutrality
constraint Eq.~(\ref{eq:6}) reflects a model degeneracy

\begin{eqnarray}
C_m\dot{V}=\frac{\omega_i}{\gamma}(\dot{\mathit{K}}_i+\dot{\mathit{Na}}_i-\dot{\mathit{Cl}}_i)\label{eq:1_new}
\end{eqnarray}
that occurs when $\textit{Na}_i$ is modelled explicitly with $\dot{\textit{Na}}_i=-\gamma/\omega_i(I_{\textit{Na}^+}-2 I_p)$
(for details see Ref.\ \cite{HUE14}). With a fixed leak current Eq.~(\ref{eq:1_new}) becomes
\begin{eqnarray}
C_m\dot{V}=\frac{\omega_i}{\gamma}(\dot{\mathit{K}}_i+\dot{\mathit{Na}}_i)-I_\mathit{leak,f}\ ,\label{eq:2_new}
\end{eqnarray}
which implies that $V=-70$ mV is a necessary fixed point condition for the system. 

In other words, the type of bistability with a second depolarized fixed point that we
normally find in closed systems is ruled out by this unphysical current. If we,
however, replace it with a chloride leak current as in our model (cf.\
Eqs.~(\ref{eq:4}) and (\ref{eq:21})), i.e., a current with a dynamically
adjusting reversal potential by virtue of Eq.~(\ref{eq:22}), we find
the same type of bistability for the closed system and monostability for the system 
that is buffered or coupled to a potassium bath. The morphological parameters
(compartmental volumes $\omega_{i/e}$ and membrane surface area $A_m$) are the same 
as for the reduced model. 

In fact in Ref.\ \cite{YAO11} the KWS model was used without additional ion regulation 
for a reaction--diffusion study of SD, and the only recovery mechanism of the local 
system seems to be this unphysical current. Theoretically SD could be a travelling wave
in a reaction--diffusion system with bistable local dynamics, but unpublished results 
show that the propagation properties in the bistable system are dramatically different 
from standard SD dynamics with wave fronts and backs travelling at different velocities. 
We hence suppose that a local potassium clearing mechanism is crucially involved in SD.

We conclude this section with a discussion of the time scales of the model. To
this end, it is helpful to keep in mind that the phenomenon of excitability
requires a separation of time scales. We have
electrical and ionic excitability and these dynamics themselves are separated
by no fewer than three orders of magnitude.

Dynamics of $V$ happens on a  scale that is faster than milliseconds.
This follows from the gating time scale $\tau_n$ which is given explicitly in
Eq.~(\ref{eq:13}) and the time scale of $\tau_V$ of $V$ which can be computed
from the membrane capacitance $C_m$ (given in Tab.~\ref{tab:1}) and the
resistance $R_m$ of the ion channels (for details see Ref.\ \cite{ERM10}):
\begin{eqnarray}
\tau_V=C_mR_m \label{eq:35}
\end{eqnarray}
with
\begin{eqnarray}
R_m = (g_{Na}^l + g_{Na}^g m^3h + g_K^l + g_K^g n^4 + g_{Cl}^l)^{-1}\ .\label{eq:36}
\end{eqnarray}
If we approximate the products of gating variables in the above expression with 0{.}1 
this gives $\tau_V\approx 0{.}07\ \mathrm{msec}$. Dynamics of $n$ happens on a scale 
in the order of milliseconds.

The time scale of ion dynamics is more 
explicit in the Goldman--Hodgkin--Katz (GHK) formalism than in the Nernst formalism 
used in this paper. The Nernst currents in Eqs.~(\ref{eq:19})--(\ref{eq:21}) are an 
approximation of the physically more accurate GHK currents, but in Ref.\ \cite{HUE14} we 
have shown that ion dynamics of GHK models and Nernst models are very similar. That is 
why the latter may be used for studies like this. For time scale considerations, 
however, we will now switch to the GHK description. The GHK current of ions with 
concentrations $\mathit{ion}_{i/e}$ across a membrane is given by
\begin{eqnarray}
I_\mathit{ion}=P_\mathit{ion}zF\xi\cdot\frac{\mathit{ion}_e\exp(-\xi)-\mathit{ion}_i}{\exp(-\xi)-1}\ , \label{eq:37}
\end{eqnarray}
where $P_{ion}$ is the permeability of the membrane to the considered ion species and 
$\xi=V/V_c$ is the dimensionless membrane potential with
\begin{eqnarray}
V_c=\frac{RT}{zF}=\frac{1}{z}\cdot 26{.}64\ \mathrm{mV}\ . \label{eq:38}
\end{eqnarray}
This expression contains the ideal gas constant $R$, the temperature $T$, ion valence 
$z$ and Faraday's constant $F$. If we now write down the GHK analogue of the ion rate 
Eqs.~(\ref{eq:3}) and (\ref{eq:4}) we obtain
\begin{eqnarray}
\frac{\mathrm{d}\mathit{ion}_i}{\mathrm{d}t}=\frac{A_m}{\omega_i}P_{ion}z\cdot\xi\cdot\frac{\mathit{ion}_e\exp(-\xi)-\mathit{ion}_i}{\exp(-\xi)-1}\ .\label{eq:39}
\end{eqnarray}
For the conversion factor $\gamma$ we have inserted the expression Eq.~(\ref{eq:11}).
The fraction term is of the order of the ion concentrations, $\xi$ is a 
dimensionless quantity and hence of order one. With the ion dynamics time scale
\begin{eqnarray}
\tau_\mathit{ion}=\frac{\omega_i}{A_mP_{ion}z}\ .\label{eq:41}
\end{eqnarray}
we can thus group the parameters as follows
\begin{eqnarray}
\frac{\mathrm{d}\mathit{ion}_i}{\mathrm{d}t}=\frac{1}{\tau_{ion}}\cdot\xi\cdot\frac{\mathit{ion}_e\exp(-\xi)-\mathit{ion}_i}{\exp(-\xi)-1}\ .\label{eq:40}
\end{eqnarray}
Permeabilities of ion channels can be found in Refs\cite{HUE14,KAG07,YAO11}. Similar as 
for the resistance $R_m$ the permeability $P_\mathit{ion}$ of a gated channel involves 
a product of gating variables. Approximating such terms again with 0{.}1 a typical 
value for the permeability is $P_\mathit{ion}\approx5\ \mu\mathrm{m}/\mathrm{sec}$. 
Together with the values for the membrane surface are and the cell volume from 
Tab.~\ref{tab:1} the time scale of transmembrane ion dynamics is 
$\tau_\mathit{ion}\approx0{.}5\ \mathrm{sec}$.

The slowest time scales are related to potassium regulation, i.e., to $\tilde{K}_e$ 
dynamics. The glia scheme from Eq.~(\ref{eq:26}) and Eq.~(\ref{eq:30}) contains a 
forward buffering process that reduces $\tilde{K}_e$ at a time scale
\begin{eqnarray}
\tau_{\textit{buff}}^\mathit{fw}=(\bar{k}_1B^0)^{-1}\label{eq:42}
\end{eqnarray}
and a backward buffering process with time scale
\begin{eqnarray}
  \tau_{\textit{buff}}^\mathit{bw}=\frac{1}{k_1}\ .\label{eq:43}
\end{eqnarray}
With the parameters from Tab.~\ref{tab:1} this leads to 
$\tau_{\textit{buff}}^\mathit{fw}\approx40\ \mathrm{sec}$ and 
$\tau_{\textit{buff}}^\mathit{bw}\approx5\ \mathrm{h}$. So backward buffering is much 
slower. This is an important property, because in the following section we will see 
that recovery from FES requires a strong reduction of the potassium content. If 
buffering and backward buffering would happen on the same time scale the required 
potassium reduction would not be possible. Backward buffering could well happen at a 
considerably faster scale than Eq.~(\ref{eq:43}), but as soon as 
$\tau_{\textit{buff}}^{fw}$ is comparable to $\tau_{\textit{buff}}^{bw}$ the buffer 
cannot re--establish physiological conditions after FES.

The glia scheme here is phenomenological. A more biophysically detailed model would 
describe a glial cell as a third compartment. An elevation of ECS potassium leads to 
glial uptake. Spatial buffering, i.e., the fast transfer of potassium ion between glia 
cells with elevated concentrations to regions of lower concentrations maintains an 
almost constant potassium concentrations in the glial cells. In SD potassium in the ECS 
is strongly elevated during an about 80 sec lasting phase of FES and is continuously 
buffered during this time. After 80 sec the concentration quickly reduces to slightly 
less than the normal physiological level. Still there is a local potassium deficit and 
what we call backward buffering, i.e., the release of potassium from the glial cells sets 
in. It is much slower than the uptake, because it is driven by a far smaller deviation 
of the potassium concentration from physiological resting conditions of the glial cell. 
So as for diffusion the forward and the backward process do not actually happen 
simultaneously.

Similar to the above explanation of slow backward buffering in the glia scheme, an 
extremely slow backward time scale follows naturally in diffusive coupling. For diffusion
the potassium content is reduced at a time scale
\begin{eqnarray}
\tau_{\textit{diff}}=\frac{1}{\lambda}\approx 35\ \mathrm{sec} \label{eq:44}
\end{eqnarray}
if extracellular potassium is greater than $K_{bath}$. Backward diffusion, however, only 
occurs in the final recovery phase that sets in after the neuron has returned from the 
transient FES state and is repolarized. While $K_i$ is still far from the resting state 
level, $K_e$ is comparable to normal physiological conditions (see the below bifurcation 
diagrams in Figs.~\ref{fig:1}b and \ref{fig:2}b) and hence the driving force
$(K_{bath}-K_e)$ during the final recovery phase is very small for a bath concentration 
close to the physiological resting state level. Consequently backward diffusion is much 
slower than forward diffusion.

Note that this argument for different slow regulation time scales only relies on the 
values of the ECS potassium concentration along the physiological fixed point branch, and 
is not a feature of the particular regulation scheme we apply.

\begin{table}[t]
\caption{\label{tab:1} Parameters for ion--based model}
\begin{tabular}{|l|l|l|}
\hline
Name 				& Value \& unit 																				& Description \\ \hline
$C_m$ 				& 1 $\mu$F/cm$^2$ 																				& membrane capacitance \\ 
$\phi$ 				& 3/msec 																						& gating time scale parameter\\ 
$g_{Na}^l$			& 0{.}0175 mS/cm$^2$ 																			& $\text{Na}^+$ leak cond. \\
$g_{Na}^g$			& 100 mS/cm$^2$ 																				& max.\ gated $\text{Na}^+$ cond. \\
$g_K^l$				& 0{.}05 mS/cm$^2$ 																				& $\text{K}^+$ leak cond. \\
$g_K^g$				& 40 mS/cm$^2$ 																					& max.\ gated $\text{K}^+$ cond. \\
$g_{Cl}^l$			& 0{.}02 mS/cm$^2$ 																				& $\text{Cl}^-$ leak cond. \\
$\mathit{Na}_i^0$	& 25{.}23 $\mathrm{mM}$																			& initial ICS $\text{Na}^+$ conc. \\
$\mathit{Na}_e^0$	& 125{.}31 $\mathrm{mM}$																		& initial ECS $\text{Na}^+$ conc. \\
$K_i^0$				& 129{.}26 $\mathrm{mM}$																		& initial ICS $\text{K}^+$ conc. \\
$K_e^0$				& 4 $\mathrm{mM}$																				& initial ECS $\text{K}^+$ conc. \\
$\mathit{Cl}_i^0$	& 9{.}9 $\mathrm{mM}$																			& initial ICS $\text{Cl}^-$ conc. \\
$\mathit{Cl}_e^0$	& 123{.}27 $\mathrm{mM}$																		& initial ECS $\text{Cl}^-$ conc. \\
$E_{Na}^0$			& 39{.}74 mV																					& initial $\text{Na}^+$ Nernst potential \\
$E_K^0$				& -92{.}94 mV 																					& initial $\text{K}^+$ Nernst potential \\
$E_{Cl}^0$			& -68 mV																						& initial $\text{Cl}^-$ Nernst potential \\
$\omega_i$ 			& 2{,}160 $\mu$m$^3$ 																			& ICS volume \\ 
$\omega_e$			& 720 $\mu$m$^3$ 																				& ECS volume \\
$F$ 				& 96485 C/mol	 																				& Faraday's constant \\ 
$A_m$ 				& 922 $\mu$m$^2$ 																				& membrane surface area \\ 
$\gamma$ 			& 9{.}556e--2 $\mu\text{m}^3\tfrac{\text{mM}}{\text{msec}}\tfrac{\text{cm}^2}{\mu\text{A}}$ 	& conversion factor \\ 
$\rho$ 				& 6{.}8 $\mu$A/cm$^2$ 																			& max.\ pump current \\
$\bar{k}_1$			& 5e--5/$\mathrm{sec}/(\mathrm{mM})$															& buffering rate \\
$k_1$				& 5e--5/sec																						& backward buffering rate \\
$\lambda$ 			& 3e--2/sec																						& diffusive coupling strength \\
$K_\mathit{bath}$	& 4 $\mathrm{mM}$																				& $\text{K}^+$ conc. of extracell.\ bath \\
$B^0$				& 500 $\mathrm{mM}$																				& initial buffer conc. \\
\hline
\end{tabular}
\end{table}

\section*{Results}\label{sec:3}

The results are presented in three parts that describe (i) the stability of closed 
models, where we treat the change $\tilde{K}_e$ of the potassium content as a bifurcation 
parameter, (ii) open models, i.e., $\tilde{K}_e$ becomes a dynamical variable, with 
glial buffering  and (iii) oscillations in ion concentrations in  open models for bath 
coupling with the bath concentration $K_\mathit{bath}$ as a bifurcation parameter.

\subsection*{Stability of closed models}
At first we will not treat the change $\tilde{K}_e$ of the potassium content as 
a dynamical variable, but as a parameter whose influence on the system's stability we 
investigate. So the model we consider is defined by the rate 
Eqs.~(\ref{eq:1})--(\ref{eq:4}) and the constraint Eqs.~(\ref{eq:5}), (\ref{eq:6}), 
(\ref{eq:8})--(\ref{eq:10}) and (\ref{eq:24}). Its stability will be important for the 
full system with dynamical ion exchange between the neuron and a bath or glial reservoir 
to be discussed in the next two subsections. The phenomenon of ionic 
excitability as in SD only occurs for dynamical 
$\tilde{K}_e$. We will see that a slow--fast decomposition of ionic excitability is 
possible. The fast ion dynamics is governed by the transmembrane dynamics that we 
discuss now and happens at the time scale 
$\tau_\mathit{ion}\approx 0{.}5\ \mathrm{sec}$. The dynamics of $\tilde{K}_e$ is much 
slower ($\tau_{\textit{buff}}^\mathit{fw}\approx40\ \mathrm{sec}$ and 
$\tau_{\textit{buff}}^\mathit{bw}\approx5\ \mathrm{h}$). Fast ion dynamics of the full 
system can hence be understood by assuming $\tilde{K}_e$ as a parameter that determines 
the level at which fast (transmembrane) ion dynamics occurs. This implies a direct 
physiological relevance of the closed system bifurcation structure with respect to potassium 
content variation for transition thresholds in the full (open) system.

\begin{figure}[t!]
\begin{center}
\includegraphics[width=0.975\columnwidth]{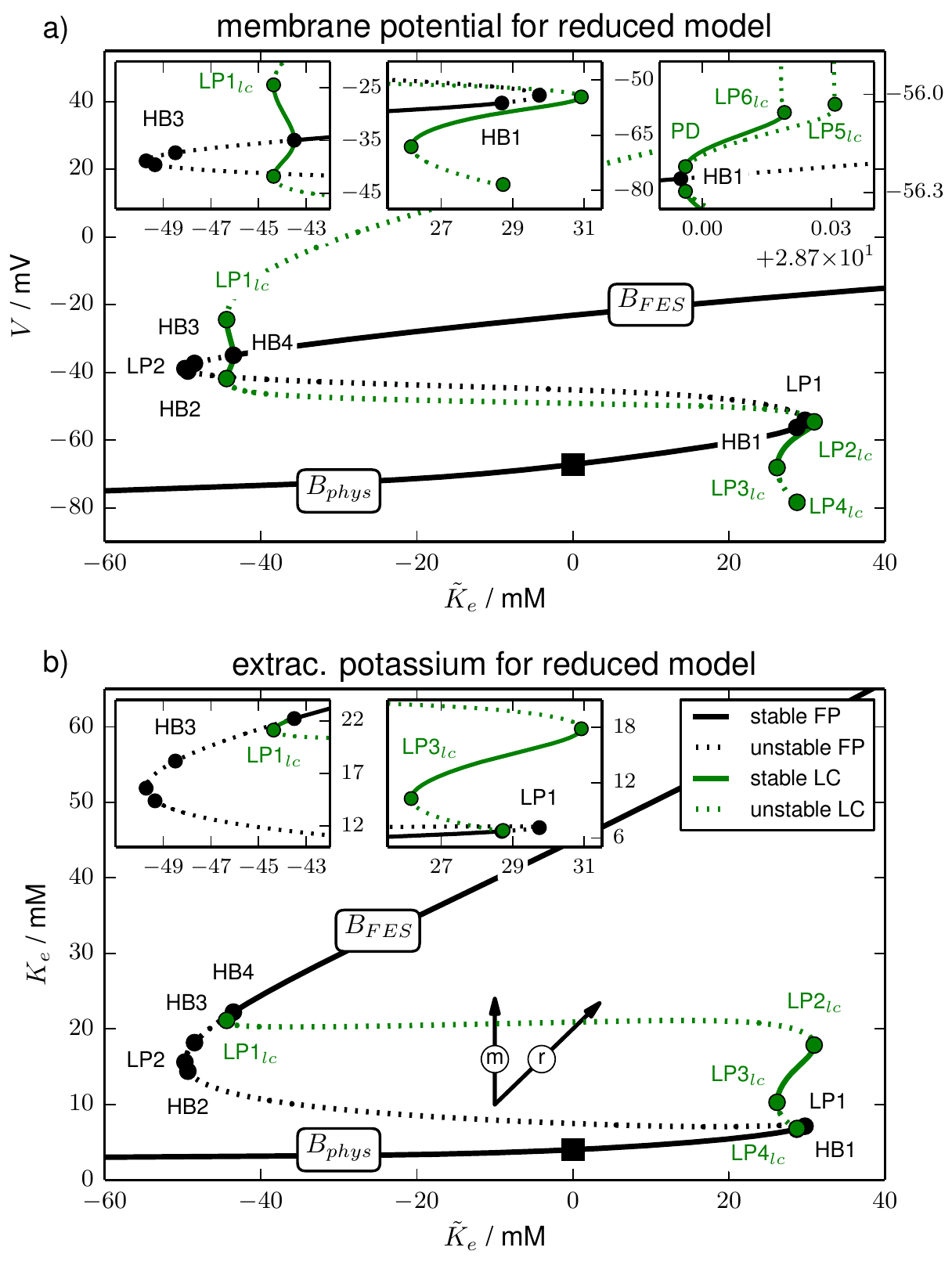}
\end{center} \caption{Bifurcation diagram of the reduced model for
$\tilde{K}_e$ as the bifurcation parameter (purely transmembrane dynamics) showing \textbf{(a)} 
the membrane potential of fixed points (FP) and limit cycles (LC), and \textbf{(b)} potassium 
concentrations. The fixed point continuation yields the black curves. Solid sections 
are fully stable, dashed sections are unstable. The stability of the fixed point 
changes in HBs and LPs. The initial physiological condition is marked by a black square. The limit 
cycle is represented by the extremal values of the dynamical variables during one 
oscillation. The continuation yields the green lines with the same stability convention 
for solid and dashed sections. The stability of the limit cycle changes either in a 
LP$_{lc}$ or in a period--doubling bifurcation (PD). In \textbf{(b)} the 
maximal and minimal extracellular potassium concentration of the limit cycle 
never differs by more than $0{.}1$ mM. The values can hence not be distinguished 
on the scale of this figure and therefore only the maximal value is drawn. The bifurcations 
are marked by full circles and labelled by the type, i.e., HB, LP or $\text{LP}_{lc}$, 
and a counter (cf.\ also the insets with blow--ups, in particular the rightmost one 
showing $\text{LP5}_{lc}$ and $\text{LP6}_{lc}$ on a very small horizontal scale). The vertical and diagonal arrows 
labelled `m' and `r' indicate the direction of extracellular potassium changes due to ion 
fluxes across the membrane (`m') and changes only due to $\tilde{K}_e$, i.e., because of 
ion exchange with a reservoir (`r'). Note that along the horizontal directions only the 
ICS potassium concentration changes by a precise mixture of fluxes across the membrane and 
ion exchange with a reservoir.  \label{fig:1}}
\end{figure}

The bifurcation diagram of the reduced model is presented in Fig.~\ref{fig:1}. It is 
shown in the $(\tilde{K}_e,V)$--plane (Fig.~\ref{fig:1}a) and in the 
$(\tilde{K}_e,K_e)$--plane (Fig.~\ref{fig:1}b) to display membrane and ion dynamics, 
respectively. A pair of arrows pointing in the direction of extracellular potassium 
changes only due to fluxes across the membrane (vertical `m' direction) and only due 
to exchange with a reservoir (diagonal `r' direction) is added to Fig.~\ref{fig:1}b.

The fixed point continuation yields a branch (black line) where fully stable 
sections are solid and unstable sections are dashed. Stability changes occur in 
saddle--node bifurcations (also called limit point bifurcation, LP) and Hopf bifurcations 
(HB). In a LP the stability changes in one direction (zero--eigenvalue bifurcation), in 
a HB it changes in two directions and a limit cycle is created (complex eigenvalue bifurcation). 
A limit cycle is usually represented by the maximal and minimal value of the dynamical 
variables. However, the oscillation amplitude of the ionic variables is almost zero for the 
limit cycles in our model. Maximal and minimal values cannot be 
distinguished on the figure scale. Hence in the $(\tilde{K}_e,K_e)$--plane the limit cycle 
continuation appears only as a single line (green). Stability changes of limit cycles occur in saddle--node 
bifurcations of limit cycles (LP$_{lc}$). The limit cycles in our model disappear in 
homoclinic bifurcations. In this bifurcation a limit cycle collides with a saddle. When 
it reaches the saddle it becomes a homoclinic cycle of infinite period.

As a reference point the initial physiological condition is marked by a black square.
We will call the entire stable fixed point branch that contains this point the physiological 
branch $B_\mathit{phys}$, because the conditions are comparable to the normal functioning 
physiological state---in particular, action potential dynamics is possible when the system is 
on this branch. 

Let us discuss the bifurcation diagram starting from this reference point and 
follow the fixed point curve in the right direction, i.e., for increasing $\tilde{K}_e$. 
The physiological fixed point loses its stability in the first (supercritical) Hopf 
bifurcation (HB1) at $\tilde{K}_e^\mathit{HB1}=28{.}7$ mM. The extracellular potassium 
concentration is then at $K_e^\mathit{HB1}=6{.}7$ mM. In other word, much of the added 
potassium has been taken up by the cell.

The limit cycle associated with HB1 
loses its stability in a period--doubling bifurcation (PD) and remains unstable. Finally 
it disappears in a homoclinic bifurcation shortly after its creation (cf.\ right inset 
in Fig.~\ref{fig:1}a). The stable limit cycle emanating from the PD point becomes 
unstable in a $\mathrm{LP}_{lc}$ and vanishes in a homoclinic bifurcation, too. The 
parameter range of these bifurcations is extremely small 
($\tilde{K}_e^\mathit{LP6_{lc}}-\tilde{K}_e^\mathit{HB1}<0{.}03\ \mathrm{mM}$). Such fine parameter 
scales will not play a role for the interpretation of ion dynamics. Ion concentrations 
are stationary and physiological up to $\tilde{K}_e^\mathit{LP6_{lc}}$, but for practical purposes it is irrelevant 
if we identify $\tilde{K}_e^\mathit{HB1}$ or $\tilde{K}_e^\mathit{LP6_{lc}}$ as the  end 
of the physiological branch $B_\mathit{phys}$.

The first HB is followed by four more bifurcations (LP1, HB2, LP2, HB3) that all neither 
restore the fixed point stability nor create any stable limit cycles. The limit cycles 
for HB2 and HB3 are hence not plotted either. It is only the fourth Hopf bifurcation (HB4) 
at $\tilde{K}_e^\mathit{HB4}=-43{.}5$ mM in which the fixed point becomes stable again 
and in which a stable limit cycle is created. The limit cycle branch loses its stability in 
LP1$_{lc}$ and regains it in LP2$_{lc}$. It becomes unstable again and even more 
unstable in LP3$_{lc}$ and LP4$_{lc}$. Shortly after that (not resolved on the scales 
in Fig.~\ref{fig:1}) it ends in a homoclinic bifurcation with the saddle between 
HB1 and LP2. At HB4 the stable free energy--starved branch $B_\mathit{FES}$ 
begins. It is generally characterized by a strong increase in the ECS potassium 
compared to physiological resting conditions (Fig.~\ref{fig:1}b), and a significant membrane 
depolarization (Fig.~\ref{fig:1}a). Corresponding to the extracellular elevation intracellular 
potassium is significantly lowered. This goes along with inverse changes of the 
compartmental sodium concentrations (all not shown). $B_\mathit{FES}$ is hence 
characterized by largely reduced ion gradients and strong membrane depolarization. In 
fact, at this membrane potential the sodium channels are inactivated which 
is normally called depolarization block in HH--like membrane models without ion dynamics. 
Depolarization block is, however, only one feature of FES. The closeness of FES to the 
thermodynamic equilibrium of the system is more importantly manifested in the reduced ion 
gradients. On $B_\mathit{FES}$ no more bifurcations occur and it remains stable for 
increasing $\tilde{K}_e$.

The interpretation of this bifurcation diagram should be as follows. The end of 
$B_\mathit{phys}$ defines the maximal potassium content compatible with a physiological 
state of a neuron. For larger $\tilde{K}_e$ it will be inevitably driven to the FES. In 
other words the end of $B_\mathit{phys}$ marks the threshold value for a slow, gradual 
elevation of the potassium content to cause the transition from physiological resting 
conditions to FES. In a buffered system it is the threshold for SD ignition. On the 
other hand stable FES--like conditions require a minimal potassium content which marks 
the end of $B_\mathit{FES}$. It is given by $\tilde{K}_e^\mathit{LP1_{lc}}=-44{.}4$ mM. Below 
this value the only stable fixed point is physiological. Again there is a narrow range, 
namely $\tilde{K}_e$ between $\tilde{K}_e^\mathit{LP1_{lc}}$ and $\tilde{K}_e^\mathit{HB4}=-43{.}5$ 
mM, in which stable oscillations can occur.

When glial buffering is at work the end of $B_\mathit{FES}$ defines the
threshold for potassium buffering, i.e., for the potassium reduction that is
required to return from FES to physiological conditions (cf.\
Eq.~(\ref{eq:25})). In the second subsection of Sect.\ Results, we will see that
this is exactly how ion regulation facilitates recovery in SD models.

There is another way the bifurcation diagram in Fig.~\ref{fig:1}b can 
be read. As we have remarked above the limit cycles of the model are characterized by 
large oscillation amplitudes in the membrane variables $n$ (not shown) and $V$, but 
almost constant ionic variables $K_{i/e}$, $\mathit{Na}_{i/e}$ and $\mathit{Cl}_{i/e}$ 
(only $K_e$ shown). So Fig.~\ref{fig:1}b tells us which extracellular 
potassium concentrations can possibly be stable and which ones cannot. Values below the 
end of $B_\mathit{phys}$ at $K_e^\mathit{HB1}=6{.}7$ mM, values between $K_e^\mathit{LP3_{lc}}=10{.}2$ 
mM and $K_e^\mathit{LP2_{lc}}=17{.}8$ mM and finally concentrations in the range 
of $B_\mathit{FES}$ starting at $K_e^\mathit{LP1_{lc}}=21{.}1$ mM can be stable. Any other 
extracellular potassium concentration is unstable and the system will evolve towards 
a stable ion configuration that is present in the phase space. The highest 
stable potassium concentration below FES values is $K_e^\mathit{LP2_{lc}}$. If potassium in
the ECS is increased instantaneously, this value indicates the threshold for SD ignition or the 
transition to FES.

\begin{figure}[t!]
\begin{center}
\includegraphics[width=0.975\linewidth]{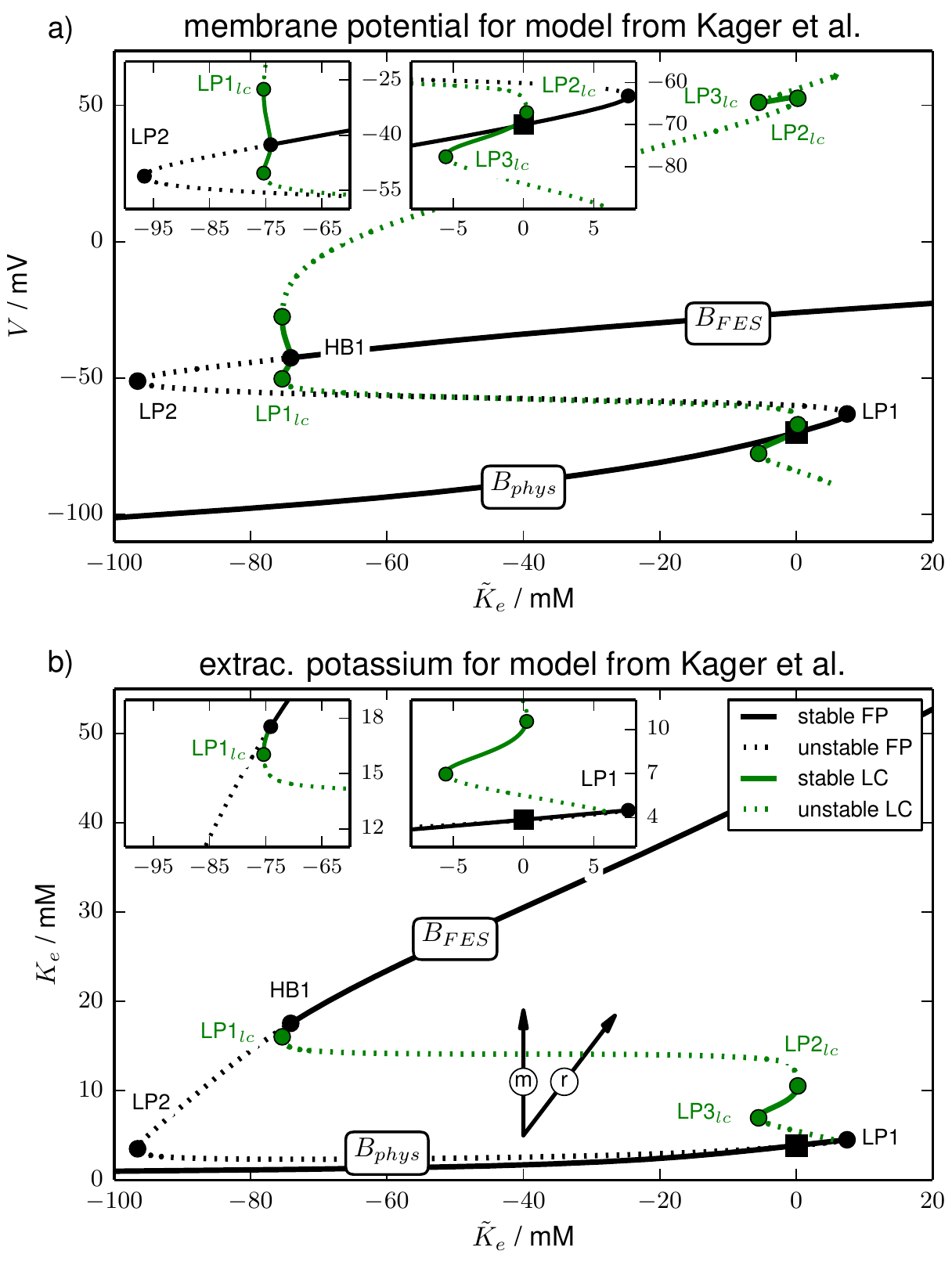}
\end{center} \caption{Bifurcation diagram of the model from Kager et al.\ (cf.\ last 
paragraph of Sect.\ Models). Like in Fig.~\ref{fig:1} panel \textbf{(a)} shows the 
membrane potential and panel \textbf{(b)} shows the extracellular potassium concentration of the 
invariant sets, i.e., fixed points and limit cycles. The line style convention (solid 
for stable, dashed for unstable) and bifurcation labels are the same as in 
Fig.~\ref{fig:1}. Note the similar shape to Fig.~\ref{fig:1}, but also the different 
scale of the two figures. \label{fig:2}}
\end{figure}

Performing the same type of bifurcation analysis with the physiologically more detailed 
model from Kager et al.\cite{KAG00,YAO11}\ (cf.\ last paragraph of Sect.\ Models) 
leads to the diagram in Fig.~\ref{fig:2}. It has been shown before that also in this 
model there is stable FES\cite{HUE14}. We do not find the same bifurcations as in the 
reduced model, but only two LPs and one HB. However, the physiological implications are 
very similar. Like in the reduced model there is an upper limit of the potassium content
$\tilde{K}_e$ for stable physiological conditions ($\tilde{K}_e^\mathit{HB1}=7{.}5$ mM) 
and a lower limit for stable FES ($\tilde{K}_e^\mathit{LP1_{lc}}=-75{.}4$ mM). Also the 
downward snaking and the stability changes of the limit cycle that starts from HB1 are
very similar to Fig.~\ref{fig:1}. This leads to the same type of conclusion concerning 
possible stable extracellular potassium concentrations. While numerical values of the 
stability limits in terms of $\tilde{K}_e$ are specific to each model, the topological 
similarity of the bifurcation diagrams suggests a generality of results: there is a 
stable physiological branch $B_\mathit{phys}$ that ends at some maximal value $\tilde{K}_e$ of 
the potassium content. Beyond this point the neuron cannot maintain physiological 
conditions, but will face FES. On the other hand the stable FES branch $B_\mathit{FES}$ ends 
for a sufficiently reduced potassium content the neuron will return to physiological 
conditions.

The new bifurcation diagrams presented in this section  confirm our 
results from Ref.\ \cite{HUE14}: Neuron models whose ionic homeostasis is only provided by 
ATPase--driven pumps, but without diffusive coupling or glial buffering, will have a highly 
unphysiological fixed point that is characterized by free energy--starvation and 
membrane depolarization. However, the here presented bifurcation diagrams contain additional 
information of great importance. Using the new bifurcation parameter $\tilde{K}_e$ crucially 
extends our results from Ref.\ \cite{HUE14} by uncovering the threshold concentrations in 
extracellular potassium concentration. These are completely novel insights.

\begin{figure}[t!]
\begin{center}
\includegraphics[width=0.975\columnwidth]{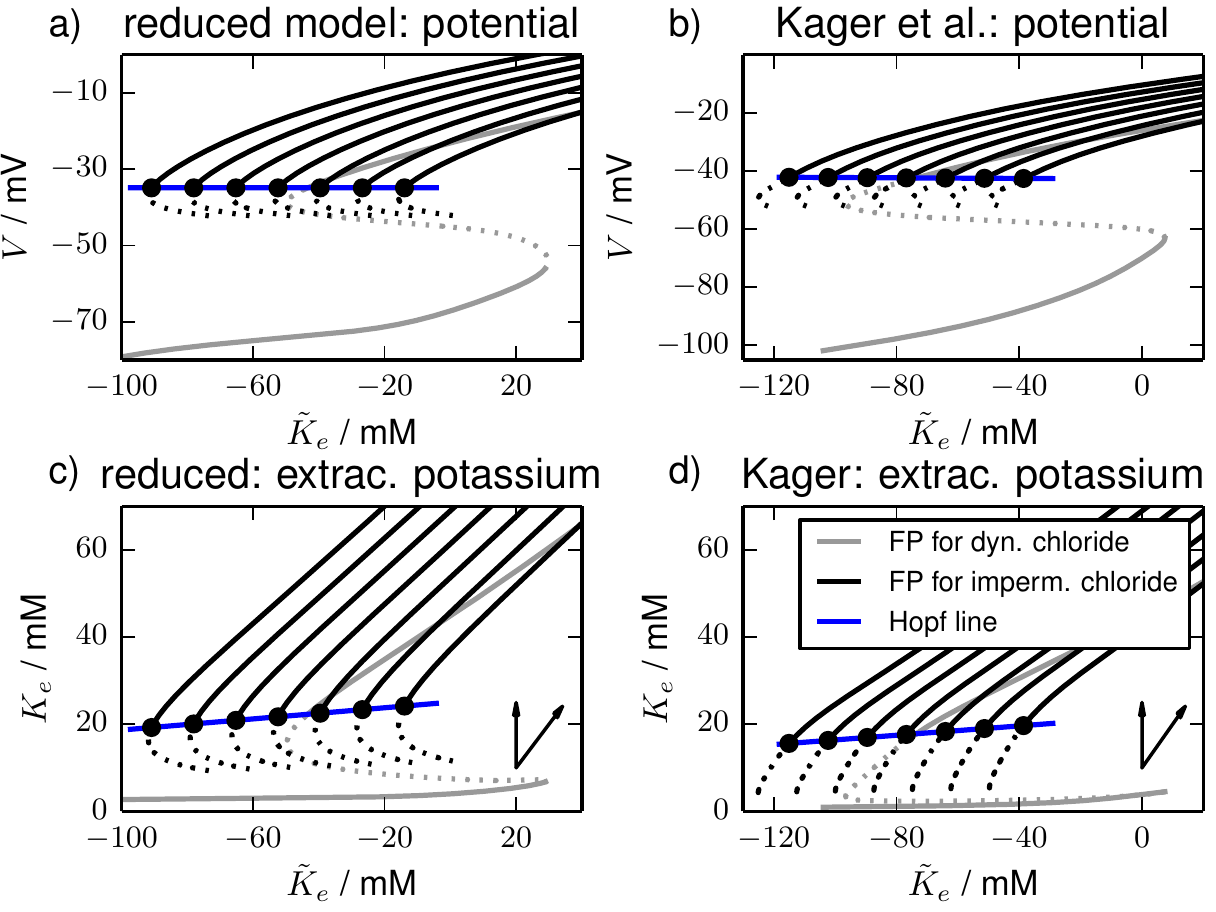}
\end{center} \caption{Fixed point continuations for a range of impermeant intracellular 
chloride concentrations $\mathit{Cl}_i$ in \textbf{(a)}, \textbf{(b)} the 
$(\tilde{K}_e,V)$--plane and \textbf{(c)}, \textbf{(d)} the $(\tilde{K}_e,K_e)$--plane. The black curves are the stable FES 
branches that lose their stability in Hopf bifurcations (black circles). Starting from 
the leftmost fixed point curves the fixed $\mathit{Cl}_i$ values are 8, 12, 16, 20, 24, 
28 and 32 mM for the reduced model and 9, 13, 17, 21, 25, 29 and 33 mM for 
the detailed model. The Hopf bifurcations for different chloride concentrations lead to 
the blue Hopf line. As a reference the fixed point curves from Figs.~\ref{fig:1} and 
\ref{fig:2} are also included to the diagram and drawn in grey. \label{fig:3}}
\end{figure}

In the next subsection the bifurcation diagrams of the unbuffered (closed)
systems shall facilitate a phase space understanding of the activation and inhibition 
process of ionic excitability as observed in SD in the buffered (open) systems. We 
are aiming for an interpretation of ionic excitability where 
neuronal discharge and recovery are fast dynamics that are governed by the bistable 
structure discussed above, whereas additional ion regulation takes the role of slowly 
changing $\tilde{K}_e$. 

However, only the gated ion dynamics, i.e., dynamics of sodium and potassium is fast 
compared to that of $\tilde{K}_e$, chloride is similarly slow. By electroneutrality this 
means that the overall concentration of positively charged ions in the ICS, i.e., the sum 
of sodium and potassium ion concentrations changes on the same slow time scale as
the chloride concentration. 

To describe this slow process not dynamically but---like $\tilde{K}_e$---in terms of a 
parameter we simply investigate the stability for a given distribution of non--dynamic, i.e., 
impermeant chloride. To determine this stability we set the chloride current to zero 
and vary $\mathit{Cl}_i$ in a certain range (from 8 to 32 mM for the reduced model, and 
from 9 to 33 mM for the detailed model). This affects the system only through the 
electroneutrality constraint Eq.~(\ref{eq:5}) which sets the intracellular charge
concentration to be shared by sodium and potassium.

For each value of $\mathit{Cl}_i$ we 
perform a fixed point continuation as in Figs.~\ref{fig:1} and \ref{fig:2} which yields 
similarly folded  s--shaped curves. The result is shown in Fig.~\ref{fig:3}. For our analysis of 
SD it is only relevant where $B_\mathit{FES}$ ends. That is why the plot does not contain the 
whole fixed point curve, but only $B_\mathit{FES}$ and a part of the unstable branch for a 
selection of $\mathit{Cl}_i$ values. As a reference the diagrams also contain the fixed 
point curves from Figs.~\ref{fig:1} and \ref{fig:2} which include chloride dynamics. 
The FES branches in Fig.~\ref{fig:3} end in Hopf bifurcations. The bifurcation points 
for different chloride concentrations yield the blue Hopf line. It marks the threshold 
for recovery from FES when dynamics of chloride and $\tilde{K}_e$ is slow.

\subsection*{Open models with glial buffering}\label{sec:4}

In the previous subsection we have analyzed the phase space structure of
ion--based neuron models without contact to a reservoir, i.e., without glial buffering
or diffusive coupling. These models have only transmembrane ion dynamics and
obey mass conservation of each ion species. Hence they describe a closed
system. The bistability of a physiological state and FES that we found in these
closed models is not experimentally observed, because real neurons are always
open systems not merely in the sense that they consume energy---a necessary
prerequisite for being far from thermodynamic equilibrium---but they also can
lose or gain ions through reservoirs or buffers. We will now include glial
buffering and show how it facilitates recovery from FES, a condition which in contrast 
to the physiological state is close to a thermodynamic equilibrium, namely the Donnan 
equilibrium (cf.\ Ref.\ \cite{HUE14}).

When glial buffering is at work, $\tilde{K}_e$ becomes a dynamical variable
whose dynamics is given by the buffering rate Eq.~(\ref{eq:30}). In
previous subsection we have explained that the bifurcation diagrams in
Figs.~\ref{fig:1} and \ref{fig:2} imply thresholds for an elevation of
extracellular potassium to trigger the transition from physiological resting conditions
to FES. This is in agreement with computational and experimental SD studies in
which high extracellular potassium concentrations are often used to trigger SD.
Another physiologically relevant way of SD ignition is the disturbance or
temporary interruption of ion pump activity. As we have shown in
Ref.\ \cite{HUE14} there is a minimal pump rate required for normal physiological
conditions in a neuron. Below this rate the neuron will go into a FES state and
remain in that state even when the pump activity is back to normal. 

\begin{figure}[t!]
\begin{center}
\includegraphics[width=0.975\columnwidth]{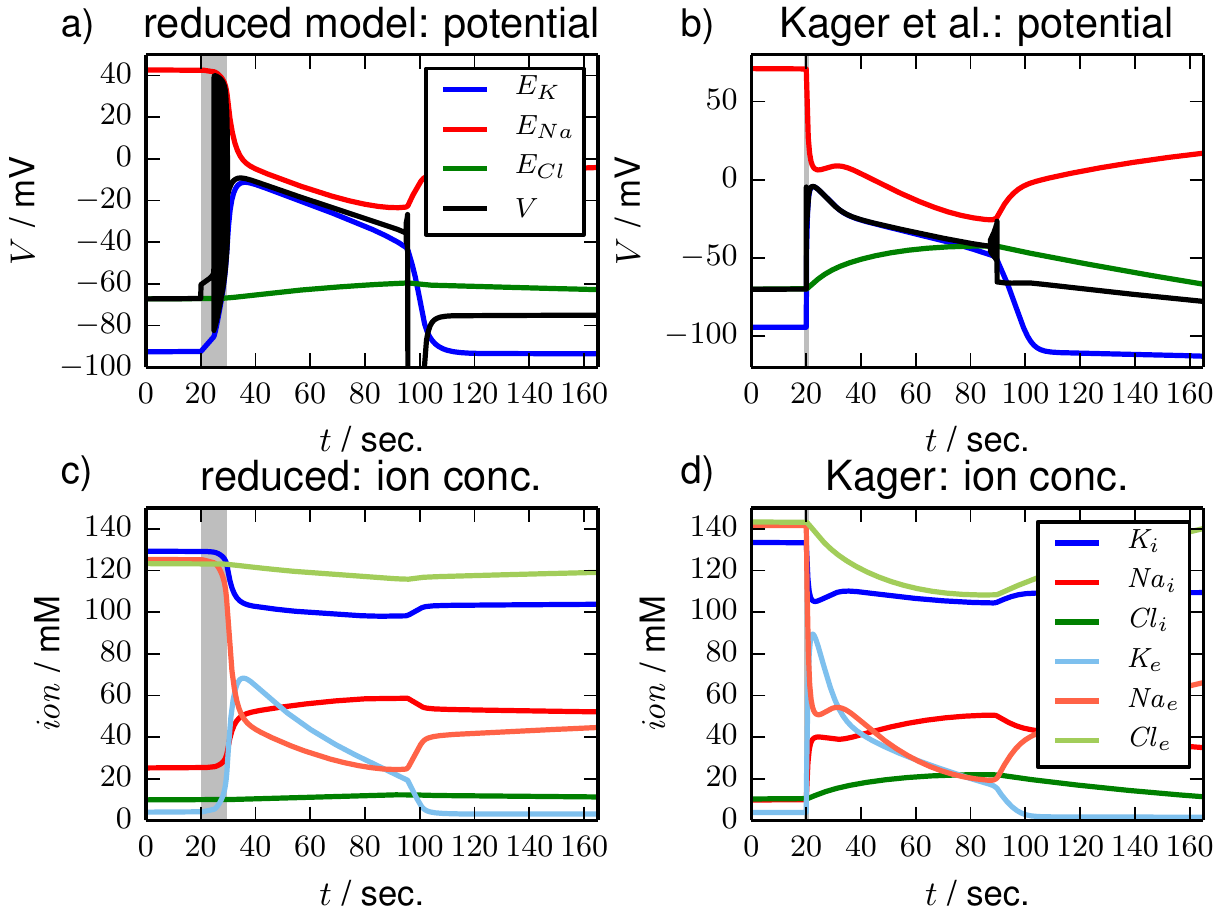}
\end{center} \caption{Time series for single SD excursions in \textbf{(a)}, \textbf{(c)}
the reduced and in \textbf{(b)}, \textbf{(d)} the detailed model. In the reduced model SD 
is triggered by an interruption of the pump activity for about 10 sec (shaded region). In the 
detailed model the extracellular potassium concentration is increased by $\Delta K_e=7{.}5$ 
mM after 20 sec (vertical line). In \textbf{(a)} and \textbf{(b)} the time series of the 
membrane potentials (black lines) are shown. Nernst potentials for all ion species are 
included to the diagrams as a reference. Ion dynamics are shown in \textbf{(c)} and \textbf{(d)} 
where extracellular ion concentrations are in lighter color. \label{fig:4}}
\end{figure}

For the simulations in Fig.~\ref{fig:4} we have interrupted the pump activity for 
about 10 sec in the reduced model, and we have elevated the extracellular potassium
concentration by $\Delta K_e=7{.}5$ mM in the detailed model to trigger
SD. Both stimulation types work for both models, but only the two examples are shown. 
The phase of pump interruption (Fig.~\ref{fig:4}a and \ref{fig:4}c) is indicated by the 
shaded region in the plots, the time of potassium elevation is marked by the vertical grey 
line. The dynamics of the two models is very similar: in response to the stimulation the 
neuron strongly depolarizes and remains in that depolarized state for about 70 sec
(Fig.~\ref{fig:4}a and \ref{fig:4}b). After that the neurons repolarize abruptly and 
asymptotically return to their initial state. In addition to the membrane potential 
(black curve) the potential plots also contain the Nernst potentials for sodium (red
line), potassium (blue line) and chloride (green line) that change with the ion
concentrations according to the definition of the Nernst potentials in
Eq.~(\ref{eq:22}). In Fig.~\ref{fig:4}c and \ref{fig:4}d we see that the potential dynamics goes
along with great changes in the ion concentrations. In particular,
extracellular potassium is strongly increased in the depolarized phase. These
conditions are very similar to the type of FES states discussed in the previous
subsection. The recovery of ion concentrations sets in with the abrupt
repolarization, but it is a very slow asymptotic process that is not shown in
Fig.~\ref{fig:4}. 

In both models the neuron is capable of producing spiking activity again right after the
repolarization. All these aspects of ion dynamics during SD are well--known
from several studies\cite{KAG00,YAO11}. We remark that the time series are
almost identical if glial buffering is replaced by the coupling to a potassium
bath. Both, the strength of glial buffering and of diffusive coupling have been adjusted
so that the depolarized phase lasts about 70 sec which is the experimentally
determined time. We will focus on bath coupling in last subsection of Sec.\ Results. 
If neither buffering nor a potassium bath is included the neuron does not repolarize 
(for time series plots of terminal transitions to FES see Ref.\ \cite{HUE14}).

\begin{figure}[t!]
\begin{center}
\includegraphics[width=0.975\columnwidth]{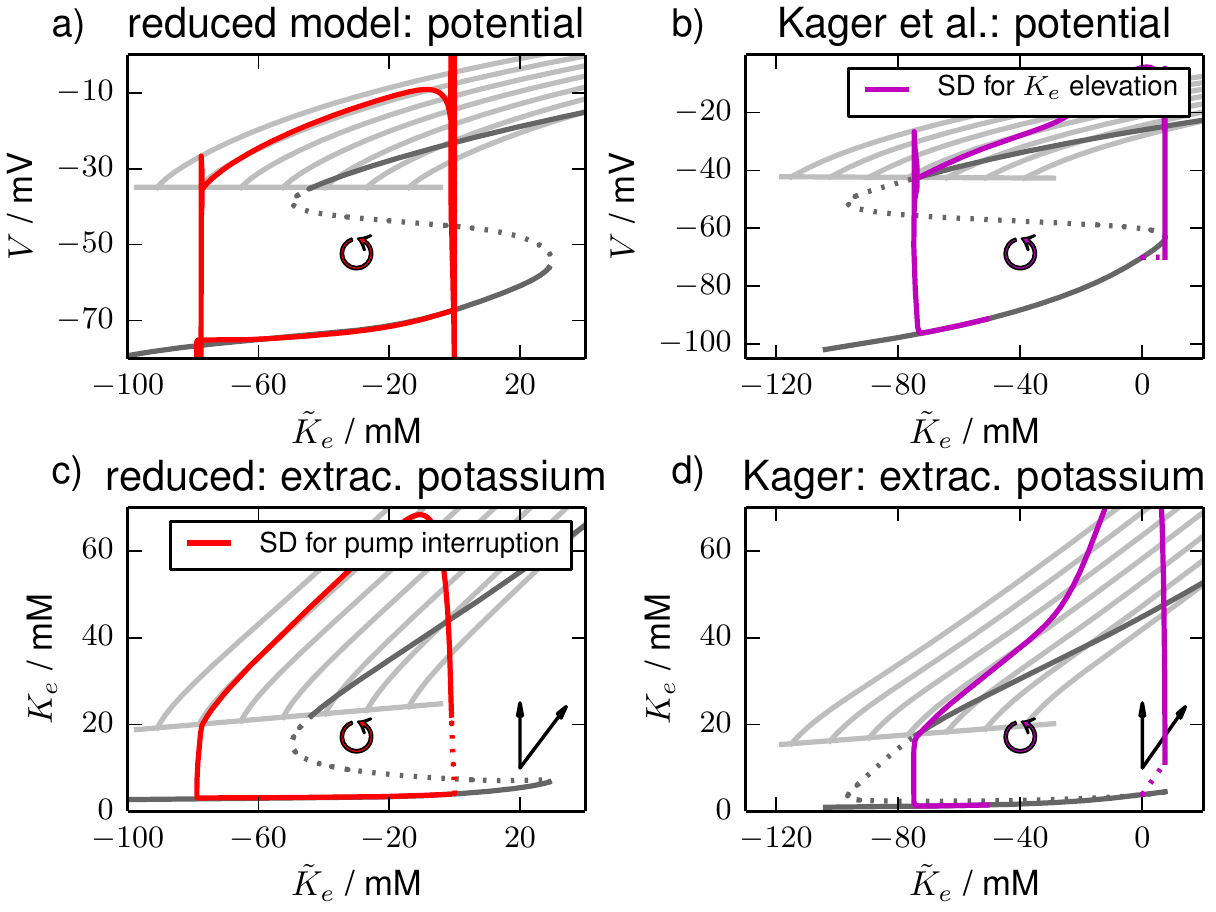}
\end{center} \caption{Phase space plots of the simulations in Fig.~\ref{fig:4}. As in 
Fig.~\ref{fig:3} panels \textbf{(a)} and \textbf{(b)} contain plots of the membrane potentials, in 
panels \textbf{(c)} and \textbf{(d)} extracellular potassium is shown. \textbf{(a)} and \textbf{(c)} 
are for the reduced model, \textbf{(b)} and \textbf{(d)} for the detailed model. The trajectories of 
the reduced model are represented as red curves, those of the detailed model are magenta. The sections 
of the trajectories that belong to times before and during the stimulation are dashed. The 
fixed point curves from Fig.~\ref{fig:3} are added to the plots as shaded lines whereas 
the fixed point continuations for the unbuffered models with dynamical chloride are 
slightly darker. The pair of arrows in the extracellular potassium plots indicates the 
direction of pure transmembrane (vertical) and pure buffering dynamics (diagonal).
\label{fig:5}}
\end{figure}

The time series in Fig.~\ref{fig:4} are useful to confirm that the neuron
models we investigate have the desired phenomenology and indeed show SD--like
dynamics. Yet the nature of the different phases of this ionic excitation
process---the fast depolarization, the prolonged FES phase and the abrupt
repolarization---remains enigmatic\cite{KAG00,KAG02,KAG07,YAO11}. In a phase
space plot the picture becomes much clearer and the entire process can be
directly related to the two stable branches, $B_\mathit{phys}$ and $B_\mathit{FES}$, that we
found for the closed and therefore pure transmembrane models in the previous
subsection. In Fig.~\ref{fig:5} the time series from Fig.~\ref{fig:4} for a
simulation time of 50 min are shown in the $(\tilde{K}_e,V)$-- and the
$(\tilde{K}_e,K_e)$--plane. The parts of the trajectories during the
stimulation (pump interruption and potassium elevation) are dashed. In the
chosen planes vertical lines belong to dynamics of constant potassium contents
that can be understood in terms of the models we analyzed in the previous subsection. That is
why Fig.~\ref{fig:5} contains the fixed point curves from Fig.~\ref{fig:3}
as shaded lines as a guide to the eye. In Fig.~\ref{fig:5}c and \ref{fig:5}d buffering 
dynamics is diagonal as indicated by the pair of arrows added to the plot. 

For both trajectories the stimulation is followed by a vertical 
activation process that leads to the transition from $B_\mathit{phys}$ to $B_\mathit{FES}$. 
The verticality means that this is a process almost purely due to transmembrane dynamics. 
It is governed by the bistable phase space structure that we discussed in the previous 
section and also in Ref.\ \cite{HUE14}. Buffering dynamics is too slow to inhibit the 
activation. The types of stimulation we applied are related to bifurcations of the 
transmembrane system: the potassium elevation is beyond the end of $B_\mathit{phys}$ which 
is marked by the first Hopf bifurcation (HB1) in Fig.~\ref{fig:1}. The interruption of 
pump activity means that we go below a pump rate threshold that is defined by a 
saddle--node bifurcation (cf.\ Ref.\ \cite{HUE14}). More generally, to initiate an 
ionic excitation it is necessary to stimulate the system until it enters the basin of 
attraction---derived in the unbuffered system---of the FES state. The activation is 
followed by a phase of both, slow transient transmembrane dynamics mostly due to 
chloride, and potassium buffering. It is the latter that bends the trajectories in the 
diagonal direction so that they go along the FES branches from Fig.~\ref{fig:3}. The 
trajectories slowly approach the repolarization threshold given by the Hopf line. The 
duration of this FES phase is determined by how long it takes the system to reach the 
Hopf line. 

This process is a mixture of buffering and transient transmembrane dynamics 
for the reduced model and more buffering--dominated in the detailed model. The duration 
of the FES phase is consequently a result of both types of dynamics. However, the main 
insight we gain from this plot is: glial buffering is the necessary inhibitory 
mechanism that takes the system to the Hopf line so that it can repolarize. We remark 
that the time series and phase space plots for bath coupling instead of buffering are 
almost identical and the same interpretation holds. The more general conclusion is then: 
ion dynamics beyond transmembrane processes is necessary to take the system to the Hopf 
line so that it can repolarize. This can, of course, be a combination of bath coupling 
and buffering. When the Hopf line is reached that neuron repolarizes abruptly which is 
the second almost purely vertical process. The repolarization is followed by slow 
asymptotic recovery dynamics of ion concentrations that takes the neuron back to the 
initial state which is at $\tilde{K}_e=0$ mM. The neuron regains the electrical 
excitability that is lost during FES already right after the repolarization. So the 
system is back to physiological function long before the ion gradients are fully 
restored.

Let us summarize the results from this subsection. By relating the SD time series from 
Fig.~\ref{fig:4} to the bifurcation structure of the unbuffered models from the first 
subsection of Sect.\ Results and in particular to the two stable branches 
$B_\mathit{phys}$ and $B_\mathit{FES}$ we have succeeded to understand ionic excitability 
as a sequence of different dynamical phases. The initial depolarization and the later 
repolarization are membrane--mediated fast processes that obey the bistable dynamics of 
unbuffered systems. The FES phase is buffering--dominated and lasts until buffering has 
taken the system to a well--defined repolarization threshold. The recovery phase is 
dominated by backward buffering. The full excursion time is the sum of the durations of 
each phase. For the de-- and repolarization process this duration mainly depends on the 
time scale of the transmembrane dynamics and is hence comparably short. The duration of 
the FES phase is a result of both, the transient transmembrane dynamics and 
glial ion regulation at a much slower time scale. The final recovery phase is mainly backward buffering 
dominated which is the slowest process. Hence the duration of an SD excursion is mainly 
determined by the slow buffering and backward buffering time scales. This conclusion that 
relies on our novel understanding of the different thresholds involved in SD is in fact in 
agreement with recent experimental data suggesting vascular clearance of extracellular 
potassium as the central recovery mechanism in SD\cite{HOF12,SUK10}.

\subsection*{Ionic oscillations for bath coupling}\label{sec:5}
The dynamics of excitable systems can often be changed to self--sustained oscillations 
by a suitable parameter variation. The type of bifurcation that leads to the 
oscillations and the shape of the limit cycle in the oscillatory regime determine 
excitation properties like threshold sharpness and latency\cite{ERM10}. The oscillatory 
dynamics that is related to ionic excitability can be obtained for bath coupling with 
an elevated bath concentration $K_\mathit{bath}$. So in this section we replace the buffering 
dynamics for $\tilde{K}_e$ with the diffusive coupling given by Eq.~(\ref{eq:33}). This 
coupling is used in experimental in--vitro studies of SD\cite{DAH03b} and has also been 
applied in computational models that are very similar to our reduced 
one\cite{CRE09,BAR11}. 

\begin{figure}[t!]
\begin{center}
\includegraphics[width=0.975\columnwidth]{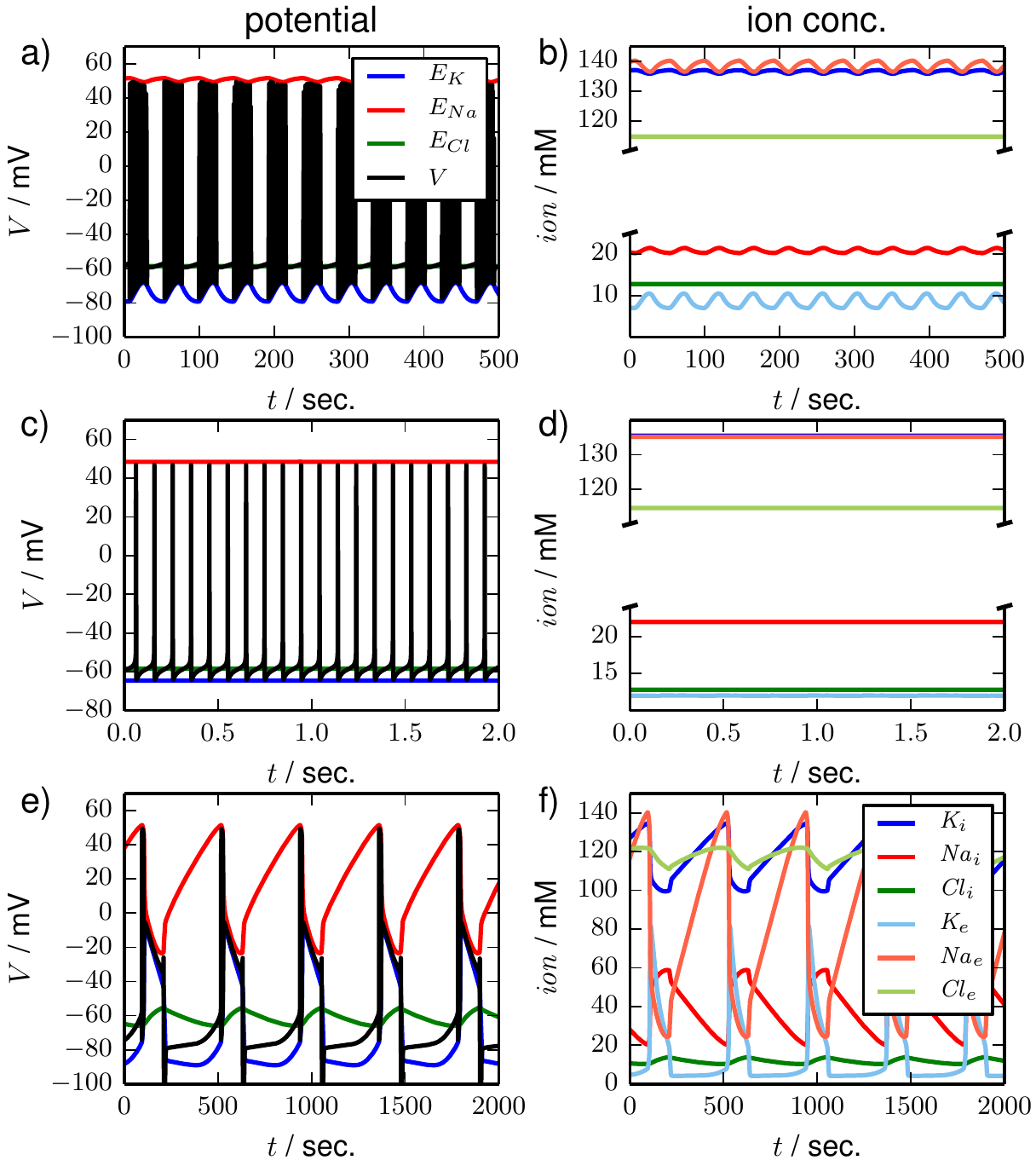}
\end{center} \caption{Time series for three types of oscillatory dynamics in the bath 
coupled reduced model. In the left panels \textbf{(a)}, \textbf{(c)} and \textbf{(e)} the 
membrane potential and the three Nernst potentials are shown. Ion concentrations are shown 
in the right panels \textbf{(b)}, \textbf{(d)} and \textbf{(f)}. The color code 
is as in Fig.~\ref{fig:4}. \textbf{(a)} and \textbf{(b)}, \textbf{(c)} and \textbf{(d)}, 
and \textbf{(e)} and \textbf{(f)} are simulations for $K_\mathit{bath}=8{.}5\ \mathrm{mM}$, 
$12\ \mathrm{mM}$ and $15\ \mathrm{mM}$, respectively. The 
dynamics is typical for \textbf{(a)} and \textbf{(b)} seizure--like activity, \textbf{(c)} 
and \textbf{(d)} tonic firing, \textbf{(e)} and \textbf{(f)} periodic SD. Note the different 
time scales of SLA, tonic firing and period SD and also the different oscillation amplitudes 
in the ionic variables. 
\label{fig:6}}
\end{figure}

Depending on the level of the bath concentration, we find three qualitatively different 
types of oscillatory dynamics that are shown in Fig.~\ref{fig:6}. The top row (a) shows 
the time series of seizure--like activity for $K_\mathit{bath}=8{.}5\ \mathrm{mM}$. It 
is characterized by repetitive bursting and low amplitude ion oscillations. The other 
types of oscillatory dynamics are tonic firing at $K_\mathit{bath}=12\ \mathrm{mM}$ with 
almost constant ion concentrations (Fig.~\ref{fig:6}b) and periodic SD at 
$K_\mathit{bath}=15\ \mathrm{mM}$ with large ionic amplitudes (Fig.~\ref{fig:6}c). We see 
that SLA and periodic SD exhibit slow oscillations of the ion concentrations and fast 
spiking activity, which hints at the toroidal nature of these dynamics. Below we will 
relate SLA and periodic SD to torus bifurcations of the tonic firing limit cycle. 

\begin{figure}[t!]
\begin{center}
\includegraphics[width=0.975\columnwidth]{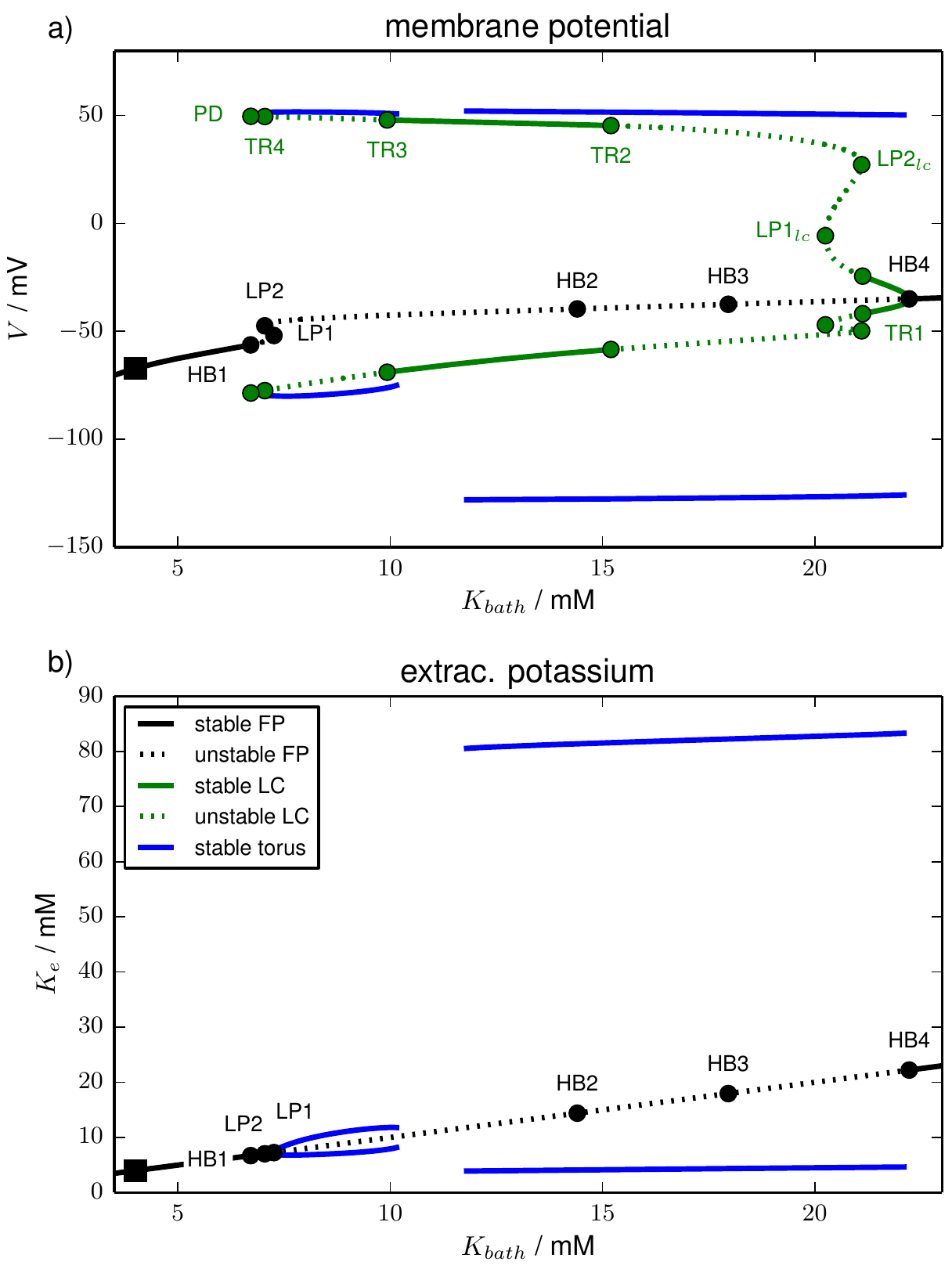}
\end{center} \caption{Bifurcation diagram of the bath coupled reduced model for 
$K_\mathit{bath}$--variation. Color and line style conventions for fixed points and limit 
cycles are Figs.~\ref{fig:1} and \ref{fig:2}: black and green lines are fixed point and 
limit cycles, solid and dashed line styles mean stable and unstable sections. Stable 
solution on invariant tori are blue. They were obtained by direct simulations. The fixed 
point changes stability in HBs and LPs. The bifurcation types limit cycle undergoes are 
$\mathrm{LP}_{lc}$, period--doubling (PD) and torus bifurcation (TR). Some 
physiologically irrelevant unstable limit cycles are omitted (cf.\ text). Panel 
\textbf{(a)} shows the membrane potential, panel \textbf{(b)} shows the extracellular potassium 
concentration. \textbf{(b)} does not contain the limit cycle, because it can hardly 
be distinguished from the fixed point line.\label{fig:7}}
\end{figure}

The examples in Fig.~\ref{fig:6} show that our model contains a variety of 
physiologically distinct and clinically important dynamical regimes. A great richness of 
oscillatory dynamics, in fact, under the simultaneous variation of $K_\textit{bath}$ and 
the glial buffering strength has already been reported in Refs.\ \cite{CRE09,BAR11} for a 
very similar model. In Ref.\ \cite{CRE09} the authors even give a bifurcation analysis of 
ionic oscillations for $K_\textit{bath}$ elevation.

To investigate dynamical changes and the transitions between the dynamical regimes in 
our model we perform a similar bifurcation analysis and vary $K_\mathit{bath}$, too. Two 
important differences should be noted though. First, Ref.\ \cite{CRE09} uses an approximation
of the multi--time scale model in which the fast spiking dynamics is averaged over time, 
while our analysis does not rely on such an approximation. Second, our analysis covers a 
bigger range of $K_\mathit{bath}$ values which allows us to compare SLA and SD, while Ref.\ 
\cite{CRE09} exclusively deals with SLA.

Fig.~\ref{fig:7} shows the bifurcation diagram for $K_\mathit{bath}$ variation in the 
$(K_\mathit{bath},V)$--plane and in the $(K_\mathit{bath},K_e)$--plane. In addition to 
fixed points (black) and limit cycles (green) also quasiperiodic torus solutions (blue) are 
contained in the diagram. In comparison to Fig.~\ref{fig:1} this model contains a new 
type of bifurcation, namely the Neimark--Sacker bifurcation, also called torus 
bifurcation (TR). A torus bifurcation is a secondary Hopf bifurcation of the radius of a 
limit cycle in which an invariant torus is created. If this torus is stable, nearby 
trajectories will be asymptotically bound to its surface. However, we cannot follow 
such solutions with standard continuation techniques, because these require an algebraic 
formulation in terms of the oscillation period. This is not possible for torus solutions, 
because on a torus the motion is quasiperiodic, i.e., characterized by two incommensurate
frequencies. We can hence only track the stable solutions by integrating the equations of 
motion and slowly varying $K_\mathit{bath}$. It is due to this numerically expensive method that 
in this section we will only analyze oscillatory dynamics of the reduced HH model with time--dependent
ion concentrations.

The result of this bifurcation analysis in Fig.\ref{fig:7} shows us that there is a maximal level 
$K_\mathit{bath}^\mathit{HB1}$ of the bath concentration compatible with physiological conditions. It 
is identified with the subcritical Hopf bifurcation HB1 in which the fixed point loses 
its stability. The related limit cycle is omitted, because it stays unstable and 
terminates in a homoclinic bifurcation with the unstable fixed point branch. The fixed 
point undergoes further bifurcations (LP1, LP2, HB2, HB3) which all leave it unstable 
and do not create stable limit cycles. It is in HB4 that the fixed point becomes 
stable again and also a stable limit cycle is created. This is the last fixed point 
bifurcation of the model.

The limit cycle that is created in HB4 changes its stability in several bifurcations. 
The physiologically most relevant ones are the four torus bifurcations. The bifurcation 
labels indicate the order of detection for the continuation that starts at HB4. 
Initially the limit cycle is characterized by fast low--amplitude oscillations. It 
becomes unstable in the subcritical torus bifurcation TR1. It regains and again loses 
its stability in the subcritical torus bifurcations TR2 and TR3. The last torus 
bifurcation, the restabilizing supercritical TR4, is directly followed by a 
PD after which no stable limit cycles exist any more. Again we have 
omitted in the diagram the unstable branch after PD and the limit cycle that is created 
in PD, which remains unstable.

Physiologically it is more intuitive to discuss the diagram for increasing
$K_\mathit{bath}$ starting from the initial physiological conditions marked by the
black square. Normal physiological conditions become unstable at $K_\mathit{bath}^\mathit{HB1}$ and
above this value the neuron spikes continuously according to the stable limit
cycle branch between PD and TR4. When $K_\mathit{bath}^\mathit{TR4}$ is reached the dynamics
changes from stationary spiking to seizure--like activity on an invariant
torus. The beginning of SLA is hence due to a supercritical torus bifurcation
and the related ionic oscillation sets in with finite period and zero
amplitude. From $K_\mathit{bath}^\mathit{TR3}$ on tonic spiking activity is stable again and
there is a small $K_\mathit{bath}$--range of bistability between SLA and this tonic
firing. As we mentioned above solutions on an invariant torus cannot be
followed with normal continuation tools like AUTO, so only stable branches are
detected.  The details of the bifurcation scenario at TR3 are hence not totally
clear, but we suspect that the unstable invariant torus that must exist near
TR3 collides with the right end of the stable torus SLA--branch in a
saddle--node bifurcation of tori. Tonic spiking then remains stable until TR2.
This bifurcation is related to the period SD that already exist well below
$K_\mathit{bath}^\mathit{TR2}$. In fact, the threshold value $K_\mathit{bath}^\mathit{TR2}$ is in 
agreement with experiments\cite{DAH03b}. Again the unstable torus near TR2 is
not detected, but we suppose that a similar scenario as in TR3 occurs. The
dynamics on the torus branch related to TR2 (and TR1 where it seems to end) is
very different from the first torus branch. While the periods of the slow
oscillations during SLA are 16--45 sec the ion oscillations of periodic SDs are
much slower with periods of 350--550 sec. 

Another crucial 
difference is obvious from Fig.~\ref{fig:7}b which shows the bifurcation 
diagram in the $(K_\mathit{bath},K_e)$--plane. The fixed point is just a straight line, because 
the diffusive coupling Eq.~(\ref{eq:33}) makes $K_e=K_\mathit{bath}$ a necessary fixed point 
condition. The limit cycle is always extremely close to this line. On the chosen scale 
it cannot be distinguished from the fixed point and is hence not contained in the plot. 
Only the torus solutions of SD and SLA attain $K_e$ values that differ significantly from 
the regulation level. The ionic amplitudes of SD are one order of magnitude larger than 
those of SLA. This has to do with the fact that the peak of SD---as described above---must 
be understood as a metastable FES state that exists due to the bistability of the 
transmembrane dynamics. The dynamics of SLA is clearly of a different nature.

Note that the bifurcation diagram reveals a bistability of tonic firing and full--blown SD 
between the left end of the SD branch at about 11 mM and TR2. This means that there is no 
gradual increase in the ionic amplitudes that slowly leads to SD, but instead it implies 
that SD is a manifest all--or--none process.

\begin{figure}[t!]
\begin{center}
\includegraphics[width=0.975\columnwidth]{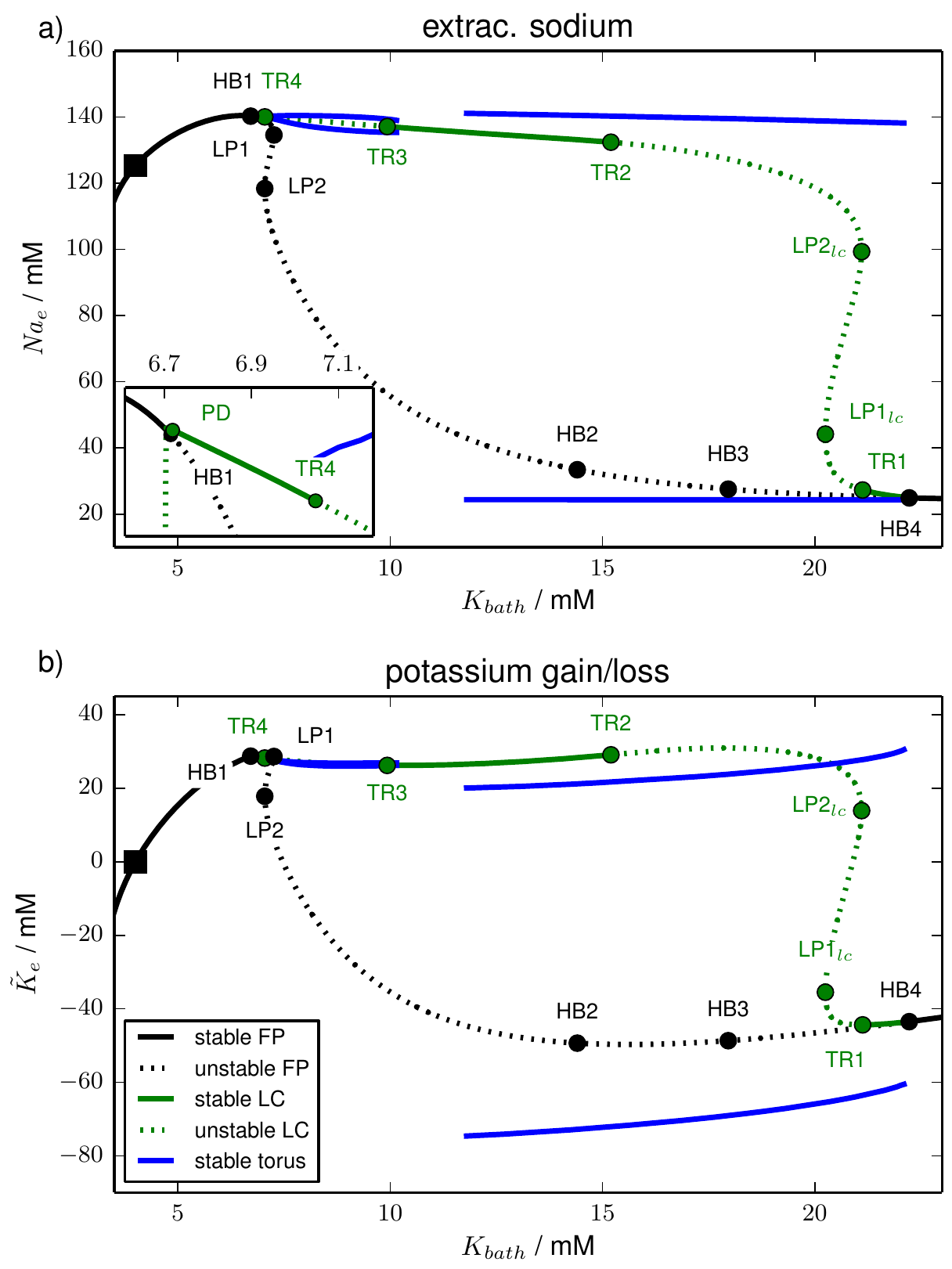}
\end{center} \caption{Different representations of the bifurcation diagram of 
Fig.~\ref{fig:7}. Panel \textbf{(a)} shows the extracellular sodium concentration and 
includes an inset around TR4 and PD. Panel \textbf{(b)} presents the potassium 
gain/loss.\label{fig:8}}
\end{figure}

In Fig.~\ref{fig:8} we look at the same bifurcation diagram in the 
$(K_\mathit{bath},\mathit{Na}_e)$-- and the $(K_\mathit{bath},\tilde{K}_e)$--plane. While in 
Fig.~\ref{fig:7} most of the ionic phase space structure is hidden, because 
$K_e\approx K_\mathit{bath}$ for fixed points and limit cycles, the 
$(K_\mathit{bath},\mathit{Na}_e)$--presentation in Fig.~\ref{fig:8}a 
provides further insights into the ion dynamics. We see that the stable fixed point 
branch before HB1 has extracellular sodium concentrations close to the physiological 
value $\mathit{Na}_e^0=125{.}31\ \mathrm{mM}$. The stable branch after HB4, however, 
has an extremely reduced extracellular sodium level and is indeed FES--like. The stable 
limit cycles between PD and TR4 and between TR3 and TR2, and also SLA are rather close 
to the physiological sodium level. On the other hand, periodic SD is an oscillation 
between FES and normal physiological conditions, which is an expected confirmation of the 
findings from the previous section. 

Fig.~\ref{fig:8}b is useful in connecting the phase space structure of 
the bath coupled system to that of the transmembrane model of the first subsection of 
Sect.\ Results. If we 
interchange the $K_\mathit{bath}$-- and the $\tilde{K}_e$--axis in the diagram it looks very 
similar to Fig.~\ref{fig:1}b. The torus bifurcations TR1, TR2 and 
TR3 are very close to the limit point bifurcations $\mathrm{LP1}_{lc}$, 
$\mathrm{LP2}_{lc}$ and $\mathrm{LP3}_{lc}$ of the transmembrane model. The fixed point 
curves are topologically identical. 

This striking similarity has to do with the fact that the limit cycle in Fig.~\ref{fig:1} 
has almost constant ion concentrations. We have pointed out in the first subsection of Sect.\ Results that 
Fig.~\ref{fig:1} tells us which extracellular potassium concentrations are stable for 
pure transmembrane dynamics. Diffusive coupling with bath concentrations at such 
potassium levels leads to negligibly small values of $J_{\textit{diff}}$ (cf.\ Eq.~(\ref{eq:33})). 
Therefore the limit cycle is still present in the bath coupled model and also the 
stability changes can be related to those in the transmembrane model. Again this can 
be seen as a confirmation of the results from the previous section: the transmembrane 
phase space plays a central role for models that are coupled to external reservoirs. 
We can interpret the ionic oscillations from Fig.~\ref{fig:6} and the bifurcations leading to 
them with respect to this phase space. 

\begin{figure}[t!]
\begin{center}
\includegraphics[width=0.975\columnwidth]{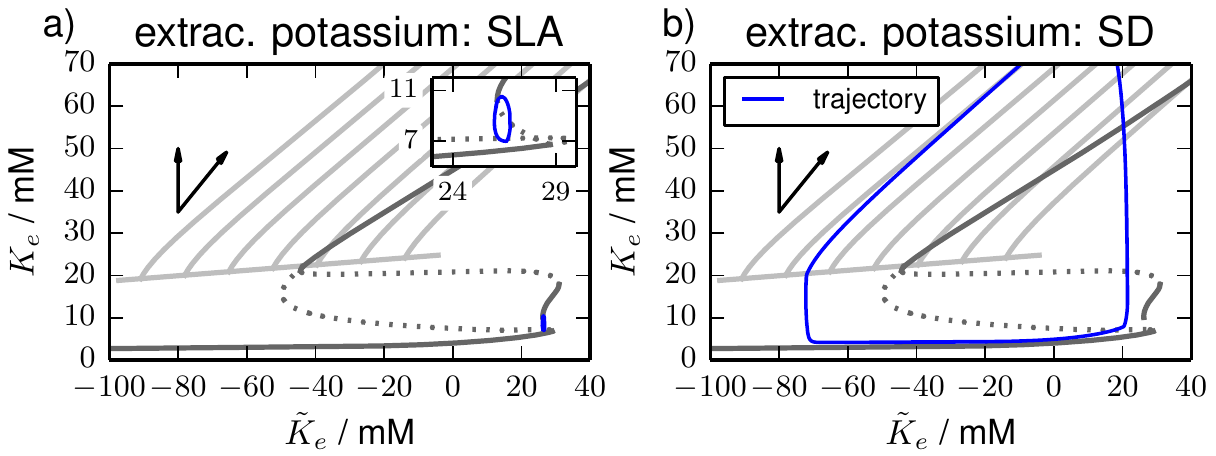}
\end{center} \caption{Phase space plots of the simulations \textbf{(a)} for SLA and \textbf{(b)}
periodic SDs from Fig.~\ref{fig:6}. Only extracellular potassium is 
shown. The limit cycle and fixed point curves from Figs.~\ref{fig:1} and \ref{fig:3} 
are superimposed to the plots as shaded lines whereas the limit cycle and fixed point from 
Fig.~\ref{fig:1} (dynamical chloride) are darker. The limit cycle and fixed point 
are not graphically distinguished, but comparison with Fig.~\ref{fig:1} should avoid 
confusion. \label{fig:9}}
\end{figure}

Last we consider the dynamics of SLA and periodic SD in a phase space projection. In 
Fig.~\ref{fig:9} the trajectories for SLA and periodic SD are plotted in the 
$(\tilde{K}_e,K_e)$--plane together with the underlying fixed point and limit cycles 
from the transmembrane model (cf.\ Fig.~\ref{fig:3}). The periodic SD trajectory has 
a very similar shape to the single SD excursion from Fig.~\ref{fig:5} and is clearly 
guided by the stable fixed point branches $B_\mathit{phys}$ and $B_\mathit{FES}$. On 
the other hand SLA is a qualitatively very different phenomenon. Rather than relating 
to the FES branch, it is an oscillation between physiological conditions and those 
stable limit cycles that exist for moderately elevated extracellular potassium 
concentrations. The ion concentrations remain far from FES. So SLA and SD are not only 
related to distinct bifurcations, though of similar toroidal nature and branching from 
the same limit cycle, but they are also located far from each other in the 
phase space. This completes our phase space analysis of local ion dynamics in open 
neuron systems

\section*{Discussion} 
In this paper we have analyzed dynamics at different time
scales in a HH model that includes time--dependent ion
concentrations. Such models are also called second generation Hodgkin--Huxley models. They
exhibit two types of excitability, electrical and ionic excitability, which are
based on fast and slow dynamics. The time scales of these types of excitability are themselves
separated by four to five orders of magnitude. The dynamics
ranges from high--frequency bursts of about 100 Hz with short interburst
periods of the order of 10 msec (Fig.~\ref{fig:6}a) to the slow periodic SD
with frequencies of about $2\cdot10^{-3}$ Hz and periods of about 7:30 min
(Fig.~\ref{fig:6}c).

The slow SD dynamics in our model is classified as ultra--slow or near--DC (direct current) 
activity and cannot normally be observed by electroencephalography (EEG) recordings, because 
of artifacts due to the resistance of the dura (thick outermost layer of the meninges that 
surrounds the brain). However, recently subdural EEG recordings provided evidence that SDs 
occur in abundance in people with structural brain damage\cite{DRE11}. Indirect evidence 
was provided already earlier by functional magnetic resonance imaging (fMRI)\cite{HAD01} 
and patient's symptom reports combined with fMRI\cite{DAH08d} that SD also occurs in migraine 
with aura\cite{CHA13a}.

The slowest dynamics that can be accurately measured by EEG, i.e., the delta band, with 
frequencies about 0{.}5 to 4 Hz, has attracted modelling approaches much more than SD, 
which was doubted to occur in human brain until the first direct measurements were reported. 
It is interesting to compare the origin of slow time scales in such delta band models to 
our slow dynamics. 

Models of the delta band essentially come in two types. On the one hand
thalamo--cortical network and mean field models of HH neurons with fixed ion
concentrations have been studied\cite{HIN13}. In this case, a slow time scale
emerges because the cells are interconnected via synaptic connections using
metabotropic receptors that are slow, because they act through second
messengers.  On the other hand, single neuron models with currents that are not
contained in HH, namely a hyperpolarization--activated depolarizing current,
$\mathit{Ca}^{2+}$--dependent sodium and potassium currents, and a persistent sodium
current, were suggested. The interplay between these currents gives rise to
oscillations at a frequency of about 2--3 Hz\cite{TIM02b}.  It is therefore
hardly surprising that these currents, in particular the persistent sodium and
the $\mathit{Ca}^{2+}$--dependent sodium and potassium currents, have also been proposed
to play an essential role in SD\cite{KAG02,SOM08}. Furthermore,  bursting as
another example of slow modulating dynamics was studied in a pure
conductance--based model with a dendritic and an axo--somatic compartment
\cite{FRO06}.

In contrast to those approaches our results show that already dynamics in a HH
framework with time--dependent ion concentrations and buffer reservoirs 
range from seconds to hours even with the original set of
voltage--gated ion currents. Time scales from milliseconds (membrane dynamics)
to seconds (ion dynamics) and even minutes to hours (ion exchange with
reservoirs) can be directly computed from the model parameters (cf.\
Sect.\ Models). The interplay of membrane dynamics, ion dynamics and
coupling to external reservoirs (glia or vasculature) naturally leads to
dynamics typical of SLA and SD. 

\begin{figure}[t!] \begin{center}
\includegraphics[width=0.975\columnwidth]{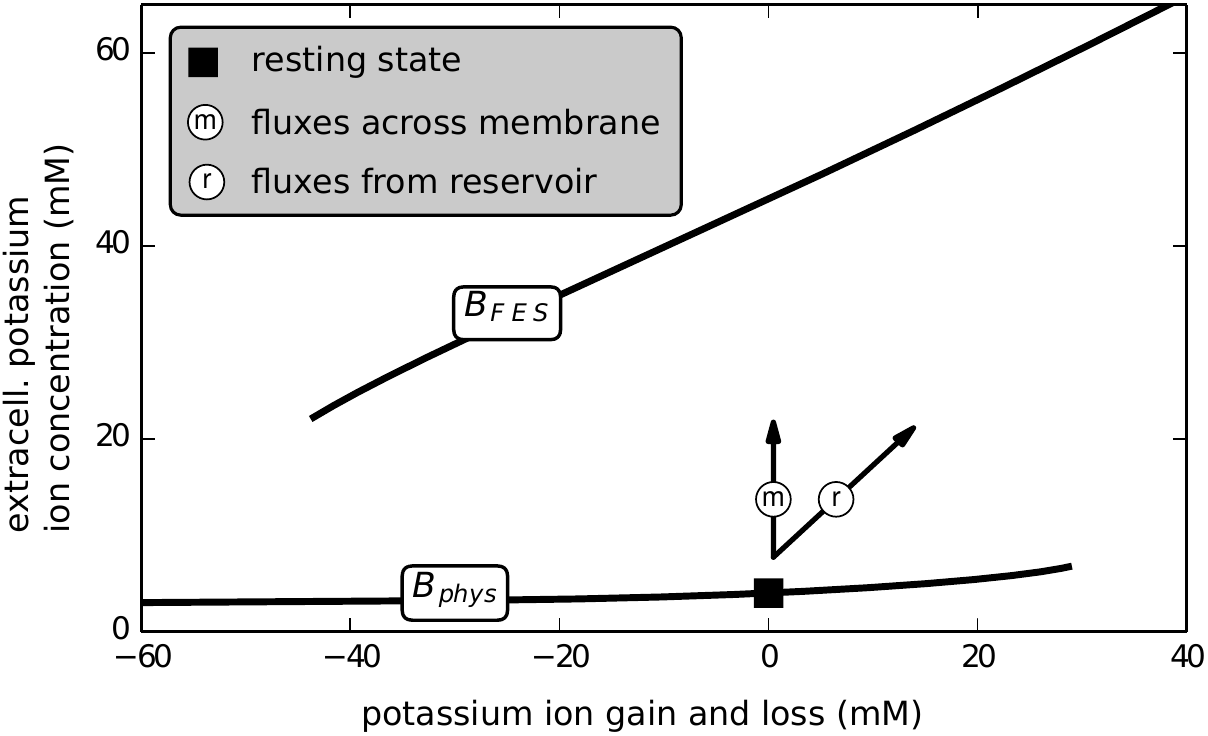} 
\end{center}
\caption{Fundamental bifurcation diagram in the slowest--scale dynamics, the
potassium ion gain or loss through reservoirs (i.e., the bifurcation
parameter).  The unit of the bifurcation parameter was chosen such that it
denotes the ion concentration with respect to the extracellular volume. The
actual extracellular potassium concentration is the order parameter. Shown
are the stable branches $B_\mathit{phys}$ and $B_\mathit{FES}$ (see Sec.\ Results) and 
the directions (arrows) of two paths of `pure'
flux condition: fluxes exclusively across the membrane and fluxes exclusively from
(or to) reservoirs.  A horizontal path is caused by a particular mixture of
these fluxes that induces potassium ion concentration changes exclusively to the
intracellular compartment. Ionic excitability can be understood as a cyclic
process in this diagram (see text).  \label{fig:bifdiagram}} 
\end{figure}

In particular SD is explained by a bistability of neuronal ion dynamics that occurs in 
the absence of external reservoirs. The potassium gain or loss $\tilde{K}_e$ through reservoirs 
provided by an extracellular bath, the vasculature or the glial cells is identified as a 
bifurcation parameter whose essential importance was not realized in earlier studies (see Fig.\ 
\ref{fig:bifdiagram}). Using this bifurcation parameter and the extracellular potassium 
concentration as the order parameter, we obtain a folded fixed point curve with the two 
outer stable branches corresponding  to states with normal physiological function, hence 
named physiological branch $B_\mathit{phys}$, and to states being free--energy starved 
($B_\mathit{FES}$). 

The definition of the bifurcation parameter implies that exchange with ion reservoirs 
happens along the diagonal direction labelled by `r'. Membrane--mediated dynamics is in the 
vertical `m' direction. In the full system where the ion exchange is a dynamical variable our 
unconventional choice of variables, i.e. modelling $\tilde{K}_e$ instead of $K_e$, makes it 
obvious that the time scales of diagonal and vertical dynamics is separated by at least two 
orders of magnitude. Slow dynamics is along $B_\mathit{phys}$ and $B_\mathit{FES}$, and the 
fast dynamics describes the jumps between these branches. We remark that dynamics along 
$B_\textit{phys}$ is slower than along $B_\textit{FES}$, because the branch is almost 
horizontal which leads to a very small gradient driving the diffusive coupling. Similarly the release 
of buffered potassium from the glia cells is only weakly driven (cf.\ the discussion of 
buffering time scales in Sect.\ Model).

In the closed system sufficiently strong stimulations lead to the transition from the physiological
resting state located on $B_\textit{phys}$ to FES. In the full system with dynamical ion exchange 
with the reservoirs, physiological conditions are restored after a large phase space excursion to the 
the before stable FES state. We refer to this process as ionic excitability. In contrast to the electrical 
excitability of the membrane potential this process involves large changes in the ion concentrations. The 
entire phase space excursion of this excitation process  can be explained through the specific transits 
between and along $B_\mathit{phys}$ and $B_\mathit{FES}$. 

We observe ion changes on three slow time scales. (i) Vertical transits between $B_\mathit{phys}$ and 
$B_\mathit{FES}$ caused by transmembrane dynamics in the order of seconds. The time scale is determined 
by the volume--surface--area ratio and the membrane permeability to the ions. (ii) Diagonal dynamics along 
$B_\mathit{FES}$ in the order of tens of seconds caused by contact to ion reservoirs. This time scale is 
determined by buffer time constants or vascular coupling strength. (iii) Dynamics on $B_\mathit{phys}$ 
again caused by contact to ion reservoirs, but at the slower backward buffering time scale in the order 
of minutes to hours determined by the slower backward rate of the buffer \cite{KAG00}. During this long 
refractory phase of ionic excitability the spiking dynamics based on electrical excitability---separated 
by seven orders of magnitude---seems fully functional.

The right end of $B_\textit{phys}$ and the left end of $B_\textit{FES}$ are marked by bifurcations that occur
for an accordingly elevated or reduced potassium content. This is the first explanation of thresholds for local 
SD dynamics in terms of bifurcations. We remark, however, that for SD ignition the important question is not
where $B_\textit{phys}$ ends, but instead where the basin of attraction of $B_\textit{FES}$ begins.

This new understanding of SD dynamics suggests a method to investigate the SD susceptibility of a given 
neuron model. One should consider the closed model without coupling to external reservoirs and check if 
shows the typical bistability between a physiological resting state and FES. We remark that unphysical 
so--called `fixed leak' currents must be replaced by proper leak currents with associated leaking ions. 
Thresholds for the transition between $B_\textit{FES}$ and $B_\textit{phys}$ translate to thresholds for SD 
ignition and repolarization, i.e., recovery from FES in the full open model. Knowledge of the potassium reduction 
needed to reach the repolarization threshold and knowledge about the buffer capacity could then tell us if 
recovery from FES can be expected (such as in migraine with aura) or if the depolarization is terminal (such 
as in stroke). 

Although our model does not contain all important processes involved in SD, our phase space explanation 
appears to be valid also for certain model extensions. For example, considering only diffusive regulation 
of potassium is physically inconsistent, but adding an analoguous regulation term for sodium turns out not 
to alter the dynamics qualitatively. Moreover osmosis--driven cell swelling---normally regarded as a 
key indicator of SD---is not included in our model, but can be added easily\cite{LEE10,SHA01,KAG02}. 
Unpublished results confirm that also with such cell swelling dynamics the fundamental bifurcation 
structure of Fig.~\ref{fig:bifdiagram} is preserved.

As a clinical application of our framework, we have linked a genetic defect, which affects the 
inactivation gate $h$ and which is present in a rare subtype of migraine with aura, to SD. 
Our simulations show that such mutations render neurons more vulnerable to SD\cite{DAH14}. 
The interesting point, however, is that on the level of the fast time scale the firing rate 
is decreased, which in a mean field approach (as done for the delta band) translates to 
decreased activity. This effect seemingly contradicts the increased SD susceptibility and 
hence illustrates the pitfalls in trying to neglect ion dynamics in the brain and to bridge 
the gap in time scales by population models.

\section*{Acknowledgement} 
The authors are grateful for discussions with Steven J.\ Schiff and Bas--Jan
Zandt. NH thanks Prof.\ Dr.\ Eckehard Sch\"oll for continuous support, fruitful
discussions, and critically reading the manuscript.  

This work was supported by the Bundesministerium f\"ur Bildung und Forschung 
(BMBF 01GQ1001B, 01GQ1109) within the Bernstein Center of Computational Neuroscience 
Berlin.


%\section*{References}
%\bibliography{ref}

\end{document}